\documentclass[aps,prb,reprint,longbibliography]{revtex4-2}%groupedaddress
\usepackage{graphicx,graphics,color,epsfig}
\usepackage{bm}
\usepackage{dcolumn}
\usepackage{amsmath}
\usepackage{amssymb}
\usepackage{graphicx}
\usepackage{epstopdf}
\usepackage[colorlinks=true,  % 启用彩色链接
    linkcolor=blue,          % 内部链接颜色（章节、图表等）
    citecolor=blue,          % 引用链接颜色
    urlcolor=blue,           % URL 链接颜色
    anchorcolor=blue]{hyperref}
\usepackage{multirow}  % 多行合并的宏包
\usepackage{lineno} % 标出行号

\newcommand{\be}{\begin{eqnarray}}
\newcommand{\ee}{\end{eqnarray}}

\newcommand{\ba}{\begin{align}}
\newcommand{\ea}{\end{align}}

\makeatletter

\newcommand{\Rmnum}[1]{\expandafter\@slowromancap\romannumeral #1@}
\makeatother

\begin{document}

\title{Kibble-Zurek scaling immune to anti-Kibble-Zurek behavior \\ in driven open systems at the limit of loss difference}
\author{Han-Chuan Kou}
\affiliation{College of Physics, Sichuan University, 610064, Chengdu, People’s Republic of China}
\author{Zhi-Han Zhang}
\affiliation{College of Physics, Sichuan University, 610064, Chengdu, People’s Republic of China}
\author{Peng Li}
\email{lipeng@scu.edu.cn}
\affiliation{College of Physics, Sichuan University, 610064, Chengdu, People’s Republic of China}

\date{\today}

\begin{abstract}
  We investigate the dissipative quench dynamics in a family of two-band fermionic systems by linearly ramping the staggered on-site energy. In the Lindblad formalism, we present an analytical solution in the presence of uniform loss or loss difference on bipartite lattices, which tells that dissipation exponentially suppresses the Kibble-Zurek (KZ) scaling behavior and the quantum jump term of the dissipation is responsible for the anti-KZ (AKZ) behavior. Interestingly, we find two different scaling behaviors at the limit of loss difference. Both scaling behaviors arise from the gapless Liouvillian. But one is accompanied by impulse stage rendered by the criticality of the system, so that it is ascribed to the universal KZ scaling law. Another depends on the dissipation strength and there is no impulse stage in it. We also point out a convenient way to observe the two new scaling behaviors by counting the number of residual particles in the end, since it is immune to the influence of AKZ behavior. We illustrate our findings through the prototypical one-dimensional Rice-Mele model first. Then, in the one-dimensional Shockley model and the two-dimensional Haldane model for Chern insulators, we show that the two scaling behaviors can appear together or separately with appropriate quench protocols.
\end{abstract}

%\pacs{05.50.+q, 75.50.Ee, 02.30.Tb}
\maketitle
%75.10.Jm  Quantized spin models, including quantum spin frustration
%05.50.+q  Lattice theory and statistics (Ising, Potts, etc.)
%03.65.Ud  Entanglement and quantum nonlocality (e.g. EPR paradox, Bell's inequalities, GHZ states, etc.)
%75.50.Ee  Antiferromagnetics

%\tableofcontents

\section{Introduction}

The concepts in equilibrium physics, such as phase transition, critical exponent, and universality, may play an important role in out-of-equilibrium situations \cite{Sachdev_2011}. As a cornerstone, the \emph{Kibble-Zurek Mechanism} (KZM) provides valuable insight into bridging the gap between nonequilibrium critical dynamics and equilibrium criticality \cite{Kibble_1976, Zurek_1985}. KZM indicates a nonadiabatic stage, which results from the critical slowing down, near a critical point \cite{Zurek_2005, Dziarmaga_2005, Polkovnikov_2005, Zurek_2006, Fischer_2007, Dziarmaga_2010, Polkovnikov_2011, kou_2022, kou_2023}. Recently, KZM has been employed as a tool to study the topological insulators through topological quantum phase transitions \cite{YangKun_2018, Lara_2020, Dutta_2020, LiFuX_2022, LiFx_2025}. Significant progress has been made in developing techniques to generate artificial magnetic fields and spin-orbit coupling for neutral atoms in optical lattices, potentially facilitating the realization and characterization of out-of-equilibrium topological insulators in cold atomic systems \cite{Cooper_2019, Liang_2023}.

Dissipation may fundamentally influence the critical behaviors of systems or significantly hinder the probing of them \cite{Breuer2002}. An intriguing phenomenon termed \emph{anti-KZ} (AKZ) behavior was observed in ferroelectric phase transitions, in which slower driving results in more defects and covers up the KZ behavior \cite{Griffin_2012}. An explanation has been explored in quantum systems subjected to a noisy control field \cite{Campo_2016, ZhuShiL_2017, Guo_2021, Suhas_2021, Campo_2022, Suhas_2023}.
For open systems, the dissipative quench dynamics can be cast into the formalism of Lindblad master equation and has been extensively explored \cite{Campo_2016, Dutta_2018, Plenio_2020, Vicari_2020, Kastner_2023, Rossini_2018, Dora_2023, Roosz_2023,Jara_2024, Ding_2024, Suzuki_2024}.

Nowadays, the impressive experimental advances have enabled the creation of highly controllable environment for studying the emergence of dissipative dynamics in superconducting resonator lattices \cite{Koch_2012, Houck_2017}, Rydberg atoms in optical lattices \cite{Bernien_2017}, and optomechanical systems \cite{Florian_2014, Favero_2017}. Furthermore, the scenario of engineered dissipation offers a broad prospect. Generally, the dynamics of open quantum systems, like the Lindblad formalism, consists of a Hermitian part, characterizing the coherent evolution of the system, and a non-Hermitian dissipator part,
accounting for loss of energy, information, and coherence into the environment. Tunable dissipation, achieved through loss difference (LD) on two hyperfine states, can facilitate the probing of the exceptional point in a corresponding non-Hermitian system \cite{Ren_2022}. On bipartite lattices, the application of LD provides a pathway for engineering tunable local flat bands \cite{Feng_2023}. More recently, theoretical and experimental investigations on the LD have been yielding significant insights \cite{JiaM_2019, Lapp_2019, Yong_2022, Yaohua_2023}.

Inspired by these improvements, we investigate the dissipative quench dynamics in two-band systems with loss difference applied to bipartite lattices. In previous works, a linear AKZ behavior was induced in weak dissipation \cite{Campo_2016, ZhuShiL_2017, Guo_2021, Suhas_2021, Campo_2022, Suhas_2023}. In our work, we demonstrate that the AKZ arises from the quantum jump term of the loss in the Lindblad formalism, and we find a new KZ scaling behavior at the limit of the loss difference (LLD), where one sublattice suffers no dissipation. This KZ scaling behavior can be observed by counting the number of residual particles in the end, since it is immune to the influence of AKZ behavior. Intriguingly, by appropriate quench protocols, another scaling behavior dependent of the dissipative strength can also be observed, which has nothing to do with criticality.

The remainder of this paper is organized as follows. In Sec. \ref{Sec-Lind}, we present the exact solution for uniform loss and analytical solution for LD. In the former, we demonstrate that the AKZ behavior is attributed to the quantum jump term of the dissipation. While in the later, we illustrate two scaling behaviors at LLD. The conclusions are exemplified by the one-dimensional Rice-Mele, Shockley, and two-dimensional Haldane models in Sec. \ref{Sec-RRM}-\ref{Sec-Hal} respectively. At last, a brief summary and discussion are given in Sec. \ref{Sec-Discu}.

\section{Lindblad equation} \label{Sec-Lind}

In this work, we focus on dissipative critical dynamics in both one- and two-dimensional two-band models on bipartite lattices. Assuming that the systems are initially prepared in the ground state far from the critical point and subjected to a linear driving in the presence of dissipation, we put the issue into the Lindblad equation \cite{Dutta_2018, Vicari_2020},
\begin{eqnarray}\label{Lindblad-RRM}
  \dot{\rho}=-i[H,\rho]+\sum_{j} \sum_{s} (L_{s,j}\rho L_{s,j}^\dagger-\frac{1}{2}\{\rho,L_{s,j}^\dagger L_{s,j}\}),
\end{eqnarray}
where the summations are over the units ($j$) and sublattices ($s=a,b$), $\rho$ denotes the density operator of the system,
\begin{eqnarray}
  H&=&H_t+H_u
\end{eqnarray}
is a fermionic Hamiltonian with the hopping energy $H_t=\sum_{j,j'}\sum_{s,s'}t_{jj'}^{ss'}c_{s,j}^\dagger c_{s',j'}$ and the on-site energy $H_u=u\sum_j(c_{a,j}^\dagger c_{a,j}-c_{b,j}^\dagger c_{b,j})$, $c_{s,j}^{(\dagger)}$ denotes the annihilation (creation) operator, and $L_{s,j}$ represents the jump operator associated with some kind of dissipations. We shall mainly elaborate on the configuration of loss dissipations by setting $L_{a,j}=\sqrt{\gamma_a}c_{a,j}$ and $L_{b,j}=\sqrt{\gamma_b}c_{b,j}$, where $\gamma_a$ and $\gamma_b$ are strengths of dissipations on $a$ and $b$ sublattices respectively. The linear driving is implemented through the on-site energy parameter,
\begin{eqnarray} \label{quenchprotocol}
  u=u(t)=u_i-\frac{t}{\tau_Q}~~~(0 \leq t \leq t_f),
\end{eqnarray}
where the $\tau_Q$ is the quench time and $t_f=(u_i-u_f)\tau_Q$.

Without loss of generality, we discuss in one dimension here. It is not hard to generalize the discussion to higher dimension. By a Fourier transition, $c_{a(b),q}=\sum_{j}\frac{e^{-iq\cdot R_j}}{\sqrt{N}}c_{a(b),j}$ where $N$ is the total number of units and $R_j$ denotes the Bravis lattices, we can reformulate the Hamiltonian in momentum space as
\begin{eqnarray} \label{sigma-H}
H&=&\sum\nolimits_{q} \textbf{d}(q)\cdot\sigma,
\end{eqnarray}
where $\sigma=(\sigma_x,\sigma_y,\sigma_z)$  and the components of the vector $\textbf{d}(q)=(d^x_q, d^y_q, d^z_q)$. The exact expressions of the components for specific models will be presented later. The diagonalized Hamiltonian reads $H = \sum_q \omega_q(\eta_{1,q}^\dagger\eta_{1,q}-\eta_{2,q}^\dagger\eta_{2,q})$, where we have the spectrum $\omega_q=[(d_q^z)^2+|\Delta_q|^2]^{1/2}$ with $\Delta_q=d_q^x+id_q^y$, and the quasiparticle operators $\eta_{1,q}= u_q c_{a,q}+v_qc_{b,q}$ and $\eta_{2,q}= -v_q^* c_{a,q}+u_q^*c_{b,q}$ with Bogoliubov coefficients $\left(u_{q}, v_{q}\right)\propto\left( \omega_{q}+d^z_q, d^x_q-id^y_q\right)$. The ground state is denoted as $|GS\rangle=\prod_q\eta_{2,q}^\dagger|0\rangle$, satisfying $c_{a,q}|GS\rangle=c_{b,q}|GS\rangle=0$.

\subsection{Uniform loss and AKZ behavior} \label{subsec-AKZ}
In the Lindblad equation in Eq. (\ref{Lindblad-RRM}), the full density matrix $\rho$ can be decomposed into sectors labelled by momentum $q$, so it can be expressed as $\rho=\prod_{q}\bigotimes\rho_{q}$. In each sector, the dimension of the density matrix $\rho_{q}$ reads $4\times4$, but it contains only 6 nonzero elements including $\rho_q^{(11)}$, $\rho_q^{(22)}$, $\rho_q^{(33)}$, $\rho_q^{(44)}$, $\rho_q^{(23)}$, and $\rho_q^{(32)}$ in the Hilbert space spanned by the base vectors $\{|0\rangle$, $c_{a,q}^\dagger|0\rangle$, $c_{b,q}^\dagger|0\rangle$, $c_{a,q}^\dagger c_{b,q}^\dagger|0\rangle\}$. Given that all the elements are worked out, we can deduce the dynamical density of excitations by
\begin{eqnarray}\label{nt}
  n(t)=\frac{1}{N}\sum_q p_q(t),
\end{eqnarray}
where the excitation probability $p_q(t)=\frac{1}{2}\text{Tr}(\rho\eta_{1,q}^\dagger\eta_{1,q})+\frac{1}{2}\text{Tr}[\rho(1-\eta_{2,q}^\dagger\eta_{2,q})]$.

Let us define a variable, $R_q=\rho_q^{(33)}-\rho_q^{(22)}$, fulfilling the equation,
\begin{eqnarray}\label{differential-Rtilde}
\frac{\text{d}\tilde{R}_q}{\text{d}x}&=&\frac{\delta}{2}+\frac{\delta^2}{4}\int_{x_0}^{x}\tilde{R}_q(x')dx' \nonumber \\
&-&4|\Delta_q|^2\int_{x_0}^{x}\tilde{R}_q(x')\cos\left(\frac{x^2-(x')^2}{\tau_Q}\right)dx',~~
\end{eqnarray}
where $x=t + x_0$, $x_0=-u_i\tau_Q$, $\tilde{R}_q=e^{\gamma t/2}R_q$, and $\delta=\gamma_a-\gamma_b$. The initial condition reads $\tilde{R}_q(0)=R_q(0)=1$, which means half filling of fermions and all of them reside on $b$ sublattice initially since the protocol demands $u_i\gg0$. After quench, the final on-site energy parameter meets $u_f\ll 0$. The solution of this equation is the key to all the 6 nonzero elements.

The Lindblad equation can be solved numerically. However, we seek for an analytical solution. Firstly, if there is no dissipation at all, the first two terms in Eq. (\ref{differential-Rtilde}) disappear. So we get a reduced equation that provides exact solution expressed by the cosine and sine Fresnel’s integrals \cite{Kholodenko_2012, Fai_2013}. For slow quench, we can arrive at the asymptotic solution (Please see details in Appendix \ref{appsec-1}),
\begin{eqnarray} \label{R}
  R_q=2 e^{-\pi\tau_Q|\Delta_q|^{2}}+2|u_{q}|^2-1~~~(\gamma=\delta=0).
\end{eqnarray}
And by Eq. (\ref{nt}), the final excitation density after quench can be worked out as $n\sim\tau_Q^{-\beta}$, which stands for the known KZ scaling law. This result is equivalent to the one expressed by parabolic cylinder function in the formalism of Laudau-Zener problem \cite{Dziarmaga_2005}. Secondly, we consider the uniform loss: $\gamma_a=\gamma_b>0$. Now that we still have $\delta=0$, the equation hasn't been changed mathematically. Thus the same form of exact solution also follows and leads to
\begin{eqnarray} \label{Rtilde}
  R_q = e^{- \bar{u}\gamma\tau_Q}(2 e^{-\pi\tau_Q |\Delta_q|^{2}}+2|u_{q}|^2-1)~(\gamma>\delta=0).~~
\end{eqnarray}

In calculating the density of excitations defined in Eq. (\ref{nt}), one would find that the variable $\Delta_q$ in Eqs. (\ref{R}) and (\ref{Rtilde}) can be substituted with the linearized form, $|\Delta_q|\propto |q-q_c|$, near its critical mode $q_c$, since only the modes in its vicinity arouse significant excitations. The value of $q_c$ depends on details of models. The density of excitations upon the completion of the quench dynamics turns out to be,
\begin{eqnarray}\label{n_case1}
  n = e^{-\bar{u}\gamma \tau_Q} A \tau_Q^{-\beta}+\frac{1}{2}\left(1-e^{-\bar{u}\gamma \tau_Q}\right),
\end{eqnarray}
where $A$ is a nonuniversal prefactor, $\bar{u}=|u_i-u_f|/2=|t_f-t_i|/2\tau_Q$, and the universal KZM exponent $\beta=d\nu/(1+z\nu)$ with $d$, $z$ and $\nu$ denoting dimensionality, dynamical and correlation length exponents. In Eq. (\ref{n_case1}), the first term tells that the KZ scaling behavior is exponentially suppressed by a factor $e^{-\bar{u}\gamma \tau_Q}$. The second term is resulted from the quantum jump part of the dissipation, as shown in Appendix \ref{app-Qjumpt}. For weaker dissipation and shorter quench time, the second term gives the linear AKZ behavior, $\sim\bar{u}\gamma \tau_Q$.

Basing on the above exact solution, we can explore more intriguing phenomena when the dissipation is delicately engineered. We shall exemplify the new phenomena by three models.

\subsection{Two kinds of scaling behaviors at LLD}\label{Sec-LiouD}

\subsubsection{Liouvillian quench dynamics \\ in the dissipative two-level system}

The Hamiltonian, Eq. (\ref{sigma-H}), can be reduced to the two-level system with different modes,
\begin{align} \label{HLauZ}
  H_q=\Delta_q\sigma^++\Delta_q^*\sigma^-+d_q^z\sigma^z.
\end{align}
Then, the dissipative quench dynamics of the two-level system can be entirely specified by looking at the Liouvillian eigenvalue problem \cite{Suzuki_2024}. The Lindblad equation can be expressed through the superoperator formalism,
\begin{align}
  \frac{d\rho_q}{dt}=\mathcal{L}_q\rho_q.
\end{align}
$\mathcal{L}_q$ is the Liouvillian superoperator acting on the $4\times4$ density matrix $\rho_q$. Here, we can reduce the $16\times 16$ Liouvillian to a $6\times 6$ form because $\rho_q$ contains only $6$ nonzero elements. Details on the vectorized form of $\rho_q$ and diagonalized form of $\mathcal{L}_q$ are provided in Appendix \ref{App_LiouS}.

The Liouvillian, $\mathcal{L}_q$, is non-Hermitian, and its instantaneous eigenvalues are given by
\begin{eqnarray}\label{Lq-lamdba}
  & \lambda_0=0,~~~ \lambda_{1,\pm}=-\frac{1}{2}\gamma\pm \frac{\sqrt{2}}{4}\sqrt{-A+\sqrt{B}},\nonumber\\
  &\lambda_{2,\pm}=-\frac{1}{2}\gamma\pm \frac{\sqrt{2}}{4}\sqrt{-A-\sqrt{B}}, ~~~\lambda_3=-\gamma,
\end{eqnarray}
where we have $A=16[(d_q^z)^2+\Delta_q^2]-\delta^2$ and $B=A^2+64(d_q^z)^2\delta^2$. We sort the eigenvalues in descending order as 
\begin{align}
0=\lambda_0\geq \lambda_{1,+}\geq \text{Re}(\lambda_{2,\pm})\geq \lambda_{1,-}\geq \lambda_3.
\end{align}
The eigenstate with zero eigenvalue, $\lambda_0$, corresponds the steady state. The rate at which the dissipative two-level system relaxes to the steady state is dictated by the eigenvalue difference $\Delta\lambda\equiv\lambda_0-\lambda_{1,+}$.
For small enough $\Delta_q$, the eigenvalue difference can be expanded as a function of $\Delta_q$
\begin{eqnarray}
  \Delta\lambda&=&\frac{\gamma}{2}- \frac{|\delta|}{2}+\frac{4|\delta|}{16(d_{q}^z)^2+\delta^2}|\Delta_q|^2. \label{lambda3}
\end{eqnarray}
Notably, in the case of the LLD, where $|\delta|=\gamma$, the parameter $d_q^z$ significantly influences the behavior of $\Delta\lambda$,
\begin{equation} \label{Liou-criP}
  \Delta\lambda=\left\{
  \begin{array}{cc}
    \frac{\gamma}{4(d_{q}^z)^2}|\Delta_q|^2 &  (|d_{q}^z|\gg 0),\\[8pt]
    \frac{4}{\gamma}|\Delta_q|^2   &    (d_{q}^z= 0).
  \end{array}
  \right.
\end{equation}
In fact, $d_q^z=0$ reflects the critical point of the many-body system, Eq. (\ref{sigma-H}). For systems far from the critical point ($|d_{q}^z|\gg 0$), $\Delta\lambda$ is proportional to $\gamma$, whereas for systems at the critical point ($d_{q}^z= 0$), $\Delta\lambda$ is inversely proportional to $\gamma$.

If the quench protocol, Eq. (\ref{quenchprotocol}), is applied to the two-level system with $d_q^z(t)\sim -t/\tau_Q$, we have
\begin{eqnarray}
  [\mathcal{L}_q(t_1), \mathcal{L}_q(t_2)] &\propto & \Delta_q. \label{commutatorLS}
\end{eqnarray}
When $\Delta_q\sim 0$, the commutator of the Liouvillian superoperators at different times is approximately zero. Consequently, the Liouvillian dynamics under the quench is given by
\begin{align}\label{LiouD_rhoq}
  \rho_q(t)\approx e^{\int_{0}^t\mathcal{L}_q(t')dt'}\rho_q(0).
\end{align}

\subsubsection{KZ and pseudo-KZ scaling behaviors at LLD} \label{LiouDquench}

We return to the discussion of the many-body system and assume, for simplicity, that the Hamiltonian has only one critical point. To avoid the influence of AKZ behavior, we count the density of residual fermion numbers,
\begin{eqnarray} \label{FerN}
\mathcal{N}&=&\frac{1}{N}\sum\nolimits_{q} \mathcal{N}_q,
\end{eqnarray}
where
\begin{eqnarray}
  \mathcal{N}_q&=&\mathcal{N}_{a,q}+\mathcal{N}_{b,q}=\text{Tr}[\rho_q (c_{a,q}^\dagger c_{a,q}+c_{b,q}^\dagger c_{b,q})]
\end{eqnarray}
represents the fermion number in the momentum space. Eq. (\ref{LiouD_rhoq}) provides the solution of density matrix in the momentum space, and Eq. (\ref{lambda3}) demonstrates that the Liouvillian spectral gap closes at LLD. Here, we rewrite $\Delta\lambda$ as a function of time at LLD,
\begin{align}
  \Delta\lambda(t)=\frac{4\gamma}{16(u_i-t/\tau_Q+c)^2+\gamma^2}|\Delta_q|^2,
\end{align}
where $c=d_q^z(t)-u(t)$ is a trivial constant that can be set to $c=0$.

If $u_i\delta <0$ and $|\delta|=\gamma$, where, in the initial system, one sublattice, predominantly occupied by fermions, suffers dissipation, the fermion number becomes zero,
\begin{align}
  ~~~~~~\mathcal{N}_q=0~~(u_i\delta <0~\text{and}~|\delta|=\gamma),
\end{align}
in the long-time evolution (i.e. $\gamma t_f\propto\gamma\tau_Q\gg 1$). The solution is trivial, so we do not further discuss this case.
However, if the opposite happens, where the initial system is good for fermions to reside on the sublattices that suffers no dissipation, $u_i\delta> 0$ and $|\delta|=\gamma$, the fermion number in the momentum space exhibits a Gaussian decay behavior in the long-time evolution, which is expressed as
\begin{align}
  ~~~~~~\mathcal{N}_q=e^{-f\tau_Q|\Delta_q|^2}  ~~(u_i\delta >0~\text{and}~|\delta|=\gamma),  \label{KZ-Nq}
\end{align}
where the factor $f$ depends on the initial and final parameters,
\begin{align}
  f=&\arctan\left(4u_i/\gamma\right)-\arctan\left(4u_f/\gamma\right)
\end{align}
and the further details are provided in Appendix \ref{App_LiouS}.

We find that the behavior of Eq. (\ref{KZ-Nq}) is similar to the excitation probability, $p_q=e^{-\pi\tau_Q|\Delta_q|^2}$, in the isolated system. Here, we can compare the quench dynamics of isolated and open two-band fermionic systems. In the isolated system, the quench dynamics is governed by the Landau-Zener transition, where the excitation density exhibits the KZ scaling behavior, $n=\int_q\frac{dq}{2\pi}p_q\sim\tau_Q^{-\beta}$. Therefore, the fermion density at final time in the open system, governed by Liouvillian dynamics of the two-level system, should be also expected to exhibit the same scaling behavior,
\begin{align}
  \mathcal{N}\propto [f\tau_Q]^{-\beta}.
\end{align}

If the quench dynamics crosses the critical point of Hamiltonian, the fermion density is given by
\begin{align} \label{N-kzscaling}
  \mathcal{N}\propto \tau_Q^{-\beta}
\end{align}
After the system crosses the critical point ($t> u_i\tau_Q$), the fermion number follows the KZ scaling behavior. The KZ scaling behavior in dissipative systems results from the change in the behavior of $\Delta\lambda$ near the critical point, resembling the impulse stage rendered by the system's criticality, as shown in Eq. (\ref{Liou-criP}).

However, if the quench dynamics does not cross the critical point of Hamiltonian, for example, $0\gg u_i\gg u_f$, the fermion density reads
\begin{align}
  \mathcal{N}
  \propto\left[\left(\frac{1}{u_f}-\frac{1}{u_i}\right)\gamma\tau_Q\right]^{-\beta}. \label{N-pkz}
\end{align}
Here, we refer to it as the pseudo-KZ (pKZ) scaling behavior, since the dynamics depends on the dissipation strength and has nothing to do with the critical dynamics.

In the following sections, we will demonstrate both scaling behaviors using the three models.

\section{Rice-mele model}\label{Sec-RRM}

The Hamiltonian of the Rice-Mele model reads
\begin{eqnarray}\label{Hami-RRM}
  H=&\sum_j \psi_j^\dagger(u\sigma_z+v\sigma_x)\psi_j+w(\psi_{j+1}^\dagger\sigma_+\psi_{j}+h.c.),~
\end{eqnarray}
where $\psi_j^\dagger=(c_{a,j}^\dagger, c_{b,j}^\dagger)$ and $\sigma_\alpha$'s are Pauli matrices. The parameters $u$, $v$ and $w$ represent the onsite energy, intracell hopping, and intercell hopping, respectively, as shown in Fig. \ref{plot-RRM-FermN}(a). In Fourier space, the Hamiltonian is reformulated as $H=\sum_q \textbf{d}(q)\cdot\sigma$, where $\sigma=(\sigma_x,\sigma_y,\sigma_z)$, and the components of $\textbf{d}(q)$ are given by $d^x_q=v+w\cos q,~d^y_q=w\sin q$ and $d^z_q=u$. We shall choose the reference energy scale such that $v=-w=1$. The critical point is $u_c=0$ with the critical mode $q_c=0$.

\subsection{KZ scaling behavior}

\begin{figure}[!t]%!htbp]
  \begin{center}
		\includegraphics[width=3.0 in,angle=0]{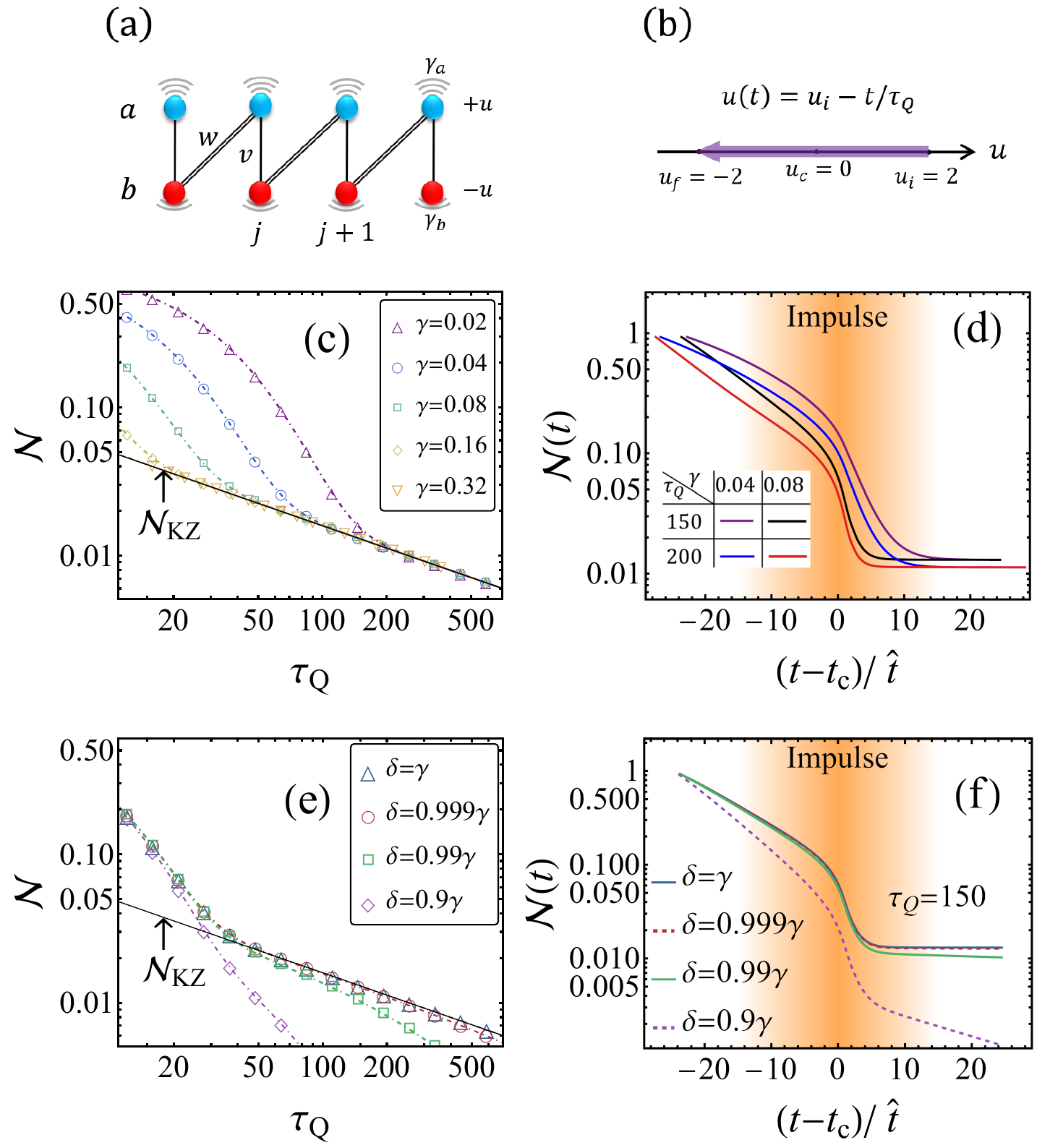}
  \end{center}
  \caption{Dissipative quench dynamics of the Rice-Mele model. (a) The model in Eq. (\ref{Hami-RRM}) in the presence of LD. (b) The quench protocol. (c) Fermion density left in the system after quench in the case of LLD ($\delta=\gamma$). The black solid line denotes the KZ scaling behavior for sufficiently long quench time. (d) Time evolution of fermion density in the case of LLD. The impulse stage near the critical point and the plateaus forming after the critical dynamics are shown. (e) and (f) Fermion density and its time evolution for several selected values of $\delta$ and $\gamma=0.08$. The dot-dashed lines in (c) and (e) are generated by the analytical formula in Eq. (\ref{RRM-FerN2}), perfectly matching the numerical solutions of the Lindblad equation indicated by the colored dingbats. In the calculation, we have set the parameters $v=-w=1$.}
  \label{plot-RRM-FermN}
\end{figure}

With the protocol shown in Fig. \ref{plot-RRM-FermN}(b), the system is ramped across the critical point. It is hard to infer exact solution from Eq. (\ref{differential-Rtilde}). Nonetheless, the exact solution presented above provides a good starting point for us to conjecture the full solution, which is given by
\begin{eqnarray}  \label{Rq_RM}
  R_q \approx & \frac{1}{e^{\bar{u}\gamma\tau_Q}}(2 e^{-\pi\tau_Q q^{2}}+2|u_{q}|^2-1) +\frac{e^{\bar{u}\delta \tau_Q}-1}{e^{\bar{u}\gamma \tau_Q}}e^{-\pi\tau_Q q^2}. \nonumber \\
\end{eqnarray}
This analytical formula agrees well with the numerical solutions of Eq. (\ref{Lindblad-RRM}) (Please see details in Appendix \ref{appsec-3}).
Subsequently, a rigorous analytical expression for the excitation density is worked out as,
\begin{eqnarray}  \label{n_case2}
  &n \approx e^{-\bar{u}\gamma \tau_Q} A \tau_Q^{-\beta}+\frac{1}{2}\left(1-e^{-\bar{u}\gamma \tau_Q}\right)+ \frac{e^{\bar{u}\delta\tau_Q}-1}{e^{\bar{u}\gamma\tau_Q}} A^{'} \tau_Q^{-\beta}.~~~~
\end{eqnarray}
For the Rice-Mele model, we have $A=1/(2\pi)$, $A^{'}=1/(4\pi)$, and $\beta=1/2$. The first two terms in Eq. (\ref{n_case2}) correspond to the exponentially suppressed KZ scaling behavior and the AKZ behavior, respectively, while the third term reflects the effect of the loss difference alone.

Now, let us concentrate on an interesting situation with LLD, i.e. $\delta=\gamma$, where the loss on one sublattice totally disappears ($\gamma_b=0$ here). For sufficiently long quench time ($\bar{u}\gamma \tau_Q\gg1$), there emerges a KZ scaling behavior,
\begin{eqnarray}\label{n_DKZ}
  \left. n-\frac{1}{2}\right|_{\delta=\gamma} \approx n_{\text{KZ}}= A^{'} \tau_Q^{-\beta}.
\end{eqnarray}
It is rigorously demonstrated through the Liouvillian dynamics, as shown in Eq. (\ref{LiouD_rhoq}). This nontrivial scaling behavior emerges after the traditional KZ scaling behavior is thoroughly suppressed by dissipation, as shown in Eq. (\ref{n_case2}). It is note worthy that we can also deduce the KZ scaling behavior for another LLD, $\delta=-\gamma$, but in this case, the protocol in Eq. (\ref{quenchprotocol}) should be reversed meanwhile.

However, the signal of KZ behavior might be submerged in the saturated value $\frac{1}{2}$, making it difficult to observe (i.e. $n_{\text{KZ}}\ll1/2$). Fortunately, we can instead detect the density of total fermion number left in the system, as given by Eq. (\ref{FerN}). It is worked out as
\begin{eqnarray}\label{RRM-FerN2}
  \mathcal{N} \approx e^{-\bar{u}\gamma\tau_Q}+\frac{e^{\bar{u}\delta\tau_Q}-1}{e^{\bar{u}\gamma\tau_Q}} A \tau_Q^{-\beta},
\end{eqnarray}
where $A=1/(2\pi)$. We see $\mathcal{N}$ does not contain any information about AKZ behavior, so we are no longer bothered by the saturated value $\frac{1}{2}$ due to dissipation any longer. Thus, by Eq. (\ref{RRM-FerN2}), the KZ behavior as shown in Eq. (\ref{N-kzscaling}) can be conveniently observed as
\begin{eqnarray}\label{N_DKZ}
  \mathcal{N}\approx\mathcal{N_\text{KZ}} = A \tau_Q^{-\beta},
\end{eqnarray}
as long as the conditions $\bar{u}\gamma \tau_Q \gg 1$ and $(\gamma-\delta)\bar{u}\tau_Q \ll 1$ are fulfilled. It is noteworthy that we now have $\mathcal{N}\approx\frac{1}{N}\sum_{q}\mathcal{N}_{b,q}\approx\mathcal{N_\text{KZ}}$, i.e. almost all remaining fermions reside on $b$ sublattices since there is no dissipation on it. As a comparison, in the absence of dissipation, we get $\mathcal{N}=1$, and the usual KZ scaling behavior should be reflected by the first term of the density of defects in Eq. (\ref{n_case2}) instead.

The KZ scaling behavior is illustrated in Fig. \ref{plot-RRM-FermN}(c). The stronger the loss dissipation on $a$ sublattice is, the shorter the quench time for the system to fall into the KZ scaling behavior.
Fig. \ref{plot-RRM-FermN}(d) illustrate the time evolution of fermion density. Similar to the KZ scaling law in the isolated system, the KZ scaling law in the dissipative system arises from the impulse stage near critical point is characterized by the frozen-out timescale $\hat{t}\sim\tau_Q^{z\nu/(1+z\nu)}$. On the other side, if we deviate from the LLD configuration, the result will acquire an exponentially decaying term, $e^{-\bar{u}(\gamma-\delta)\tau_Q}$, according to Eq. (\ref{RRM-FerN2}). For small deviation, the KZ-like behavior can still be discerned somehow as shown in Fig. \ref{plot-RRM-FermN}(e) and (f). While for large deviation ($\gamma\gg\delta$), the excitation density exponentially approaches the saturated value $\frac{1}{2}$ for strong dissipation and long quench time, because the system rapidly evolves into the pure steady state $|\psi(t_f)\rangle=|0\rangle$ and nothing interesting is left.

It is noteworthy that the KZ scaling law disclosed above can also be observed at the limit of gain difference. This can be seen by replacing the jump operators with $L_{a,j}=\sqrt{\gamma_a}c_{a,j}^{\dagger}$ and $L_{b,j}=\sqrt{\gamma_b}c_{b,j}^{\dagger}$. In this case, the system will gain lots of fermions, and the scaling behaviors would be restated by the density of the hole number instead.

\begin{figure}[t]%!htbp]
  \begin{center}
  		\includegraphics[width=3.3 in,angle=0]{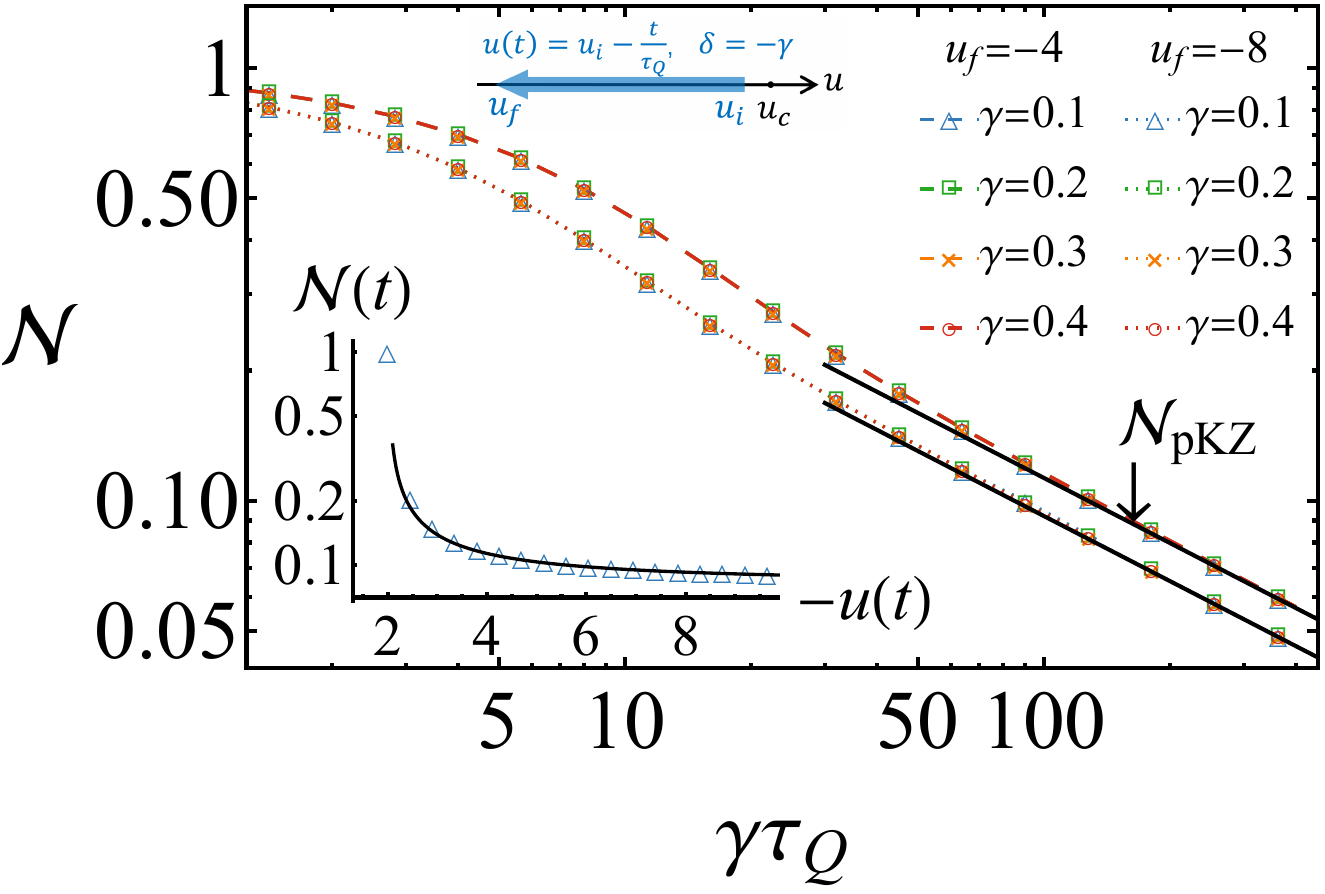}
  \end{center}
  \caption{The fermion density as a function of quench time in the case of LLD ($\delta=-\gamma$). We adopt the quench protocol $u(t)=u_i-t/\tau_Q$, with $u_i=-2$. The numerical solutions of the Lindblad equation, represented by the colored dingbats, collapse onto the pKZ scaling behavior shown in Eq. (\ref{RRM-PKZ}), represented by the black solid line. In the inset, the black solid line, generated by Eq. (\ref{RRM-PKZ-v2}), perfectly matches the numerical solutions, demonstrating that the time-dependent fermion density tends to a non-zero constant as $u(t)\rightarrow-\infty$. }
  \label{plot-RRM-PKZ}
\end{figure}

\subsection{The pseudo-KZ scaling behavior}

When $u_iu_f>0$, $u_i\delta>0$ and $|\delta|=\gamma$, the fermion density conform to pKZ scaling behavior described in Eq. (\ref{N-pkz}),
\begin{align} \label{RRM-PKZ}
  \mathcal{N}\approx \mathcal{N}_\text{pKZ}=\frac{1}{\sqrt{\frac{1}{u_f}-\frac{1}{u_i}}}\frac{1}{\sqrt{\pi\gamma\tau_Q}}.
\end{align}
The condition $u_iu_f>0$ ensures that the time-dependent system avoids crossing the critical point, and the conditions, $u_i\delta>0$ and $|\delta|=\gamma$, ensure that the initial condition is good for the fermions to reside on the sublattices that suffers no dissipation. Fig. \ref{plot-RRM-PKZ} illustrates the pKZ scaling behavior for several selected $u_f$, where we adopt the quench protocol $u(t)=u_i-t/\tau_Q$ in the case of LLD ($\gamma=-\delta$).

We can express the time-dependent fermion density by replacing $u_f$ in Eq. (\ref{RRM-PKZ}) with the parameter $u(t)$ from Eq. (\ref{quenchprotocol}),
\begin{align} \label{RRM-PKZ-v2}
  \mathcal{N}(t)\approx\frac{1}{\sqrt{\frac{1}{u(t)}-\frac{1}{u_i}}}\frac{1}{\sqrt{\pi\gamma\tau_Q}}.
\end{align}
Unlike the KZ scaling behavior, the pKZ behavior lacks the impulse stage, as shown in the inset of Fig. \ref{plot-RRM-PKZ}. Moreover, the pKZ scaling behavior approaches to a non-zero constant, $\mathcal{N}(\infty)=\sqrt{|u_i|}/\sqrt{\pi\gamma\tau_Q}$, when the parameter $u(t)$ becomes sufficiently large.

\section{Shockley model: two types of scaling behaviors}\label{Sec-Sch}

\begin{figure}[t]%!htbp]
  \begin{center}
		\includegraphics[width=3.3 in,angle=0]{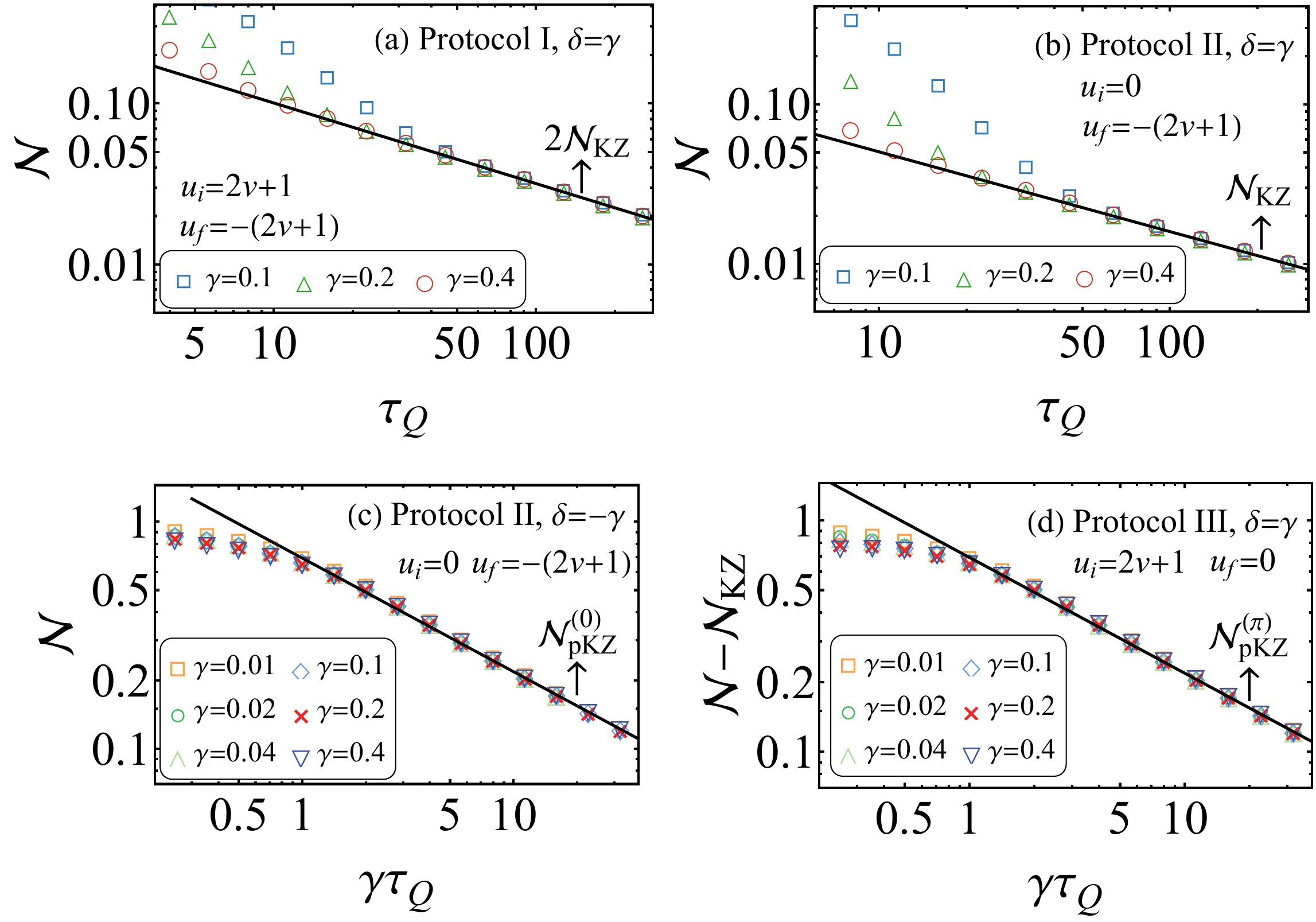}
  \end{center}
  \caption{ Dissipative quench dynamics of the Shockley model. (a)-(d) The fermion density left in the system after quench for the four cases that produce scaling behaviors, $\mathcal{N}_\text{KZ}$ and/or $\mathcal{N}_\text{pKZ}^{(0/\pi)}$, as listed in Table \ref{table-1}. In (a)-(d), we have set $v=w=1/2$ and $u_c=2v=1$.}
  \label{plot-ShocM-FermN}
\end{figure}

The Shockley model reads \cite{Edward_2023}
\begin{eqnarray}\label{Hami-ShocM}
  H= \sum_j u\psi_j^\dagger\sigma_z\psi_j-\{\psi_j^\dagger(v\sigma_z+iw\sigma_y)\psi_{j+1}+h.c.\}.~
\end{eqnarray}
Likewise, the Hamiltonian in the Fourier space can be reformulated as $H=\sum_q \textbf{d}(q)\cdot \sigma$, where $d^x_q=0,~d^y_q=2w\sin q,~d^z_q=u-2v\cos q$.
Here, we ensure $w,v\geq 0$ and consider topological quantum phase transitions. There are two critical points, $u=u_c$ and $u=-u_c$ where $u_c=2v$. The energy gap of Hamiltonian vanishes at $u=u_c$ with $q=0$, or at $u=-u_c$ with $q=\pi$. The topologically trivial and nontrivial phases are distinguished by the winding number \cite{Asboth_2016}, $W=(1/2\pi)\int_{-\pi}^{\pi}(\tilde{\textbf{d}}\times \frac{\text{d}}{\text{d}q} \tilde{\textbf{d}})_x \text{d}q$, where $\tilde{\textbf{d}}=\textbf{d}(q)/|\textbf{d}(q)|$. It takes the values $W=0$ for the trivial phase if $|u|>u_c$, and $W=1$ for the nontrivial phase if $|u|<u_c$.

When the condition of the LLD is applied to the Shockley model, $\Delta\lambda$, defined in Eq. (\ref{lambda3}), vanishes at $q_c=0$ or $q_c=\pi$. Consequently, the final fermion density is given by the integral of $\mathcal{N}_q$ near $q_c=0$ and $q_c=\pi$
\begin{align}
  \mathcal{N}=\mathcal{N}^{(0)}+\mathcal{N}^{(\pi)},
\end{align}
where $\mathcal{N}^{(q_c)}=\int_{q\sim q_c}\frac{dq}{2\pi}\mathcal{N}_q$. For the different $q_c$, the time-dependent parameter $d_q(t)$ can be simplified as
\begin{align}
  d_q(t)\rightarrow\left\{
  \begin{array}{cc}
    d_0^z(t)=u(t)-2v&(q_c=0),\\
    d_\pi^z(t)=u(t)+2v&(q_c=\pi).
  \end{array}
  \right.
\end{align}
Therefore, the $\mathcal{N}^{0/\pi}$ is given by
\begin{align}
  \mathcal{N}^{(0/\pi)}=\left\{
  \begin{array}{cc}
   \mathcal{N}_\text{KZ} & (d_{0/\pi}^z(t_i)d_{0/\pi}^z(t_f)<0),  \\[4.5pt]
   \mathcal{N}^{(0/\pi)}_\text{pKZ} & (d_{0/\pi}^z(t_i)d_{0/\pi}^z(t_f)>0),
  \end{array}
  \right.
\end{align}
where $\mathcal{N}^{(0/\pi)}_\text{pKZ}$ is the pseudo-KZ sclaing behavior,
\begin{align}
  \mathcal{N}^{(0/\pi)}_\text{pKZ}=\frac{1}{\sqrt{\frac{1}{d_{0/\pi}^z(t_f)}-\frac{1}{d_{0/\pi}^z(t_i)}}}\frac{1}{\sqrt{\pi\gamma\tau_Q}}.
\end{align}
Notice that we should ensure that the initial system is good for the fermions to reside on the sublattices that suffers no dissipation, i.e. $d_{0/\pi}^z(t_i)\delta>0$ and $|\delta|=\gamma$. As listed in Table \ref{table-1} and illustrated in Fig. \ref{plot-ShocM-FermN}, we may observe the two scaling behaviors, $\mathcal{N}_\text{KZ}$ and $\mathcal{N}_\text{pKZ}$, together or separately by the appropriate protocols. If $d_{0/\pi}^z(t_i)\delta<0$ and $|\delta|=\gamma$, we have $\mathcal{N}^{(0/\pi)}=0$ at the final time.

As an example, let us examine the case of protocol III with the LLD, $\delta=\gamma$ or $\gamma_a>\gamma_b=0$, which produces the KZ and pKZ scaling behaviors, as illustrated in Fig. \ref{plot-ShocM-FermN}(d). In this case, we have $\mathcal{N}^{(0)}=\mathcal{N}_\text{KZ}$, derived from the conditions $d_{0}^z(t_i)\delta>0$ and $d_{0}^z(t_i)d_{0}^z(t_f)<0$. Similarly, we have $\mathcal{N}^{(\pi)}=\mathcal{N}^{(\pi)}_\text{pKZ}$, derived from the conditions $d_{\pi}^z(t_i)\delta>0$ and $d_{\pi}^z(t_i)d_{\pi}^z(t_f)>0$. Therefore, the final fermion density is given by $\mathcal{N}=\mathcal{N}_\text{KZ}+\mathcal{N}^{(\pi)}_\text{pKZ}\propto \tau_Q^{-1/2}$.

\section{Haldane model}\label{Sec-Hal}

\begin{figure}[t]%!htbp]
  \begin{center}
		\includegraphics[width=3.3 in,angle=0]{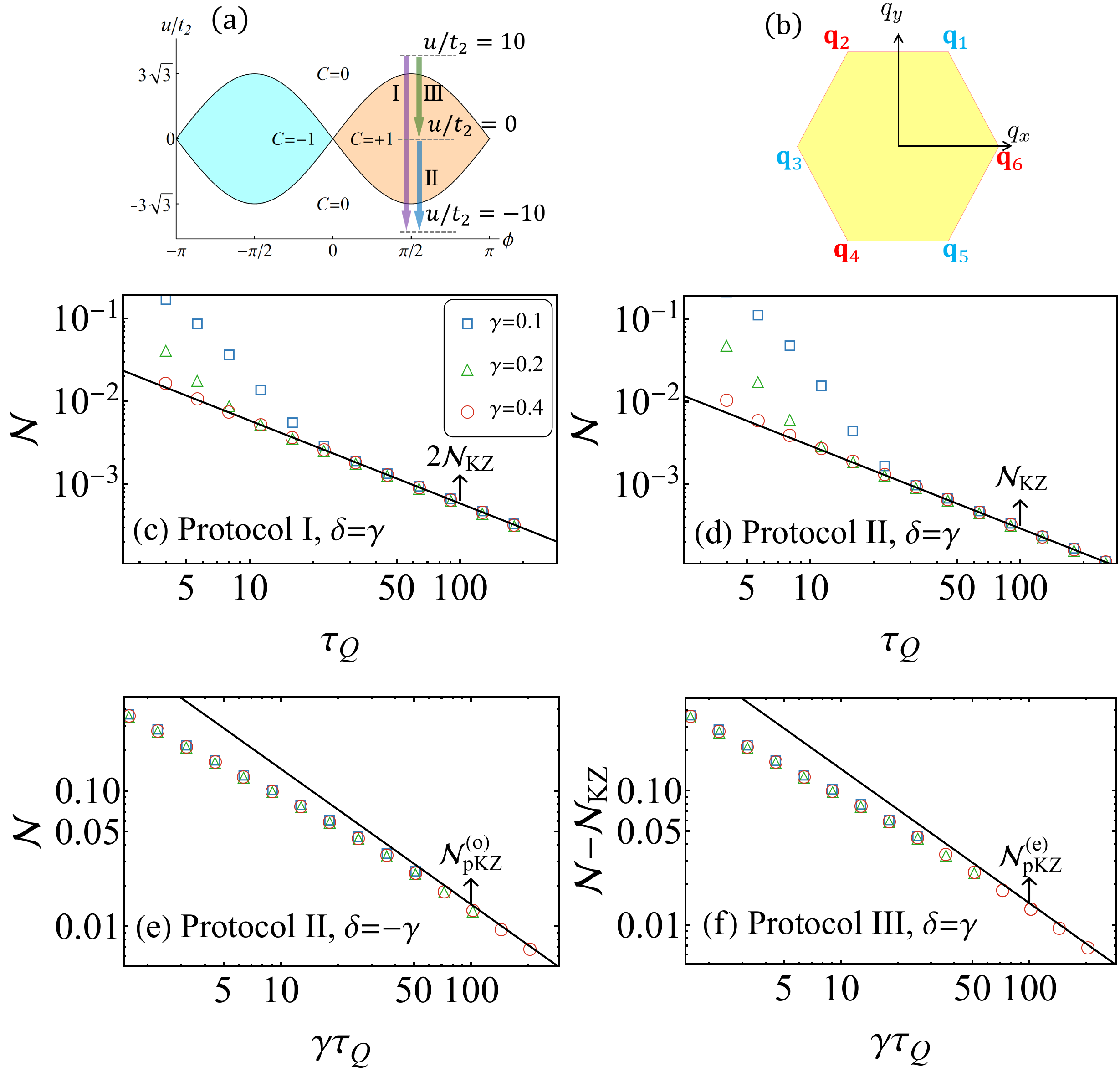}
  \end{center}
  \caption{(a) Phase diagram of the Haldane model and the quench protocols I-III. (b) First brillouin zone of Haldane model, where $\textbf{q}_i$'s ($i=1,2,\cdots,6$) are six corners. For $u=3\sqrt{3}t_2$ and $\textbf{q}_c=\textbf{q}_{1/3/5}$, or $u=-3\sqrt{3}t_2$ and $\textbf{q}_c=\textbf{q}_{2/4/6}$, the energy gap closes.  (c)-(f) Fermion density for several values of $\gamma$. The calculations are performed on $140\times140$ lattices with fixed parameters, $t_1=1$ and $t_2=1/2$. The black lines in (c) and (d) are given by the KZ scaling behavior, while those in (e) and (f) correspond to pKZ scaling behaviors.}
  \label{plot-Haldane-FermN}
\end{figure}

The Haldane model is defined on a honeycomb lattices that can be decomposed into $a$ and $b$ sublattices \cite{Haldane_1987, Shen_2017}, on which two species of fermions reside respectively. So it allows us to apply the same scenario to explore the KZ and pKZ scaling behaviors on it. The Hamiltonian is given by
\begin{align}
H=&-t_1\sum_{\langle i,j\rangle}\left(c_i^\dagger c_j+h.c.\right)
-t_2\sum_{\langle\langle i,j\rangle\rangle}\left(e^{i\phi_{ij}}c_i^\dagger c_j+h.c.\right) \nonumber\\
&+u \sum_{i\in a}c_i^\dagger c_i-u\sum_{i\in b}c_i^\dagger c_i,
\end{align}
where $t_1$ and $t_2$ is the nearest-neighbor and next-nearest-neighbor hopping term, $u$ is the on-site energy parameter, and $c_{i}^{(\dagger)}=c_{a(b),\textbf{r}}^{(\dagger)}$ is the creation (annihilation) operator on the sublattice $i\in a(b)$. $\phi_{ij}=\phi$ is the phase factor between the $a$ sublattice and $\phi_{ij}=-\phi$ between $b$ sublattice. The three basis vectors in the real space, $\textbf{a}_1=(0,1)$,  $\textbf{a}_2=(\frac{\sqrt{3}}{2},-\frac{1}{2})$ and  $\textbf{a}_3=(-\frac{\sqrt{3}}{2},-\frac{1}{2})$, represent the displacement from $b$ sublattice to its three adjacent $a$ sublattice on $x-y$ plane, where the lattice constant is $1$. The Chern number reads $C=\pm1$ for $u/t_2<3\sqrt{3}|\sin\phi|$ and $C=0$ for other situations. By fixing the parameters, $t_1=1$ and $t_2=1/2$, we obtain the phase diagram illustrated in Fig. \ref{plot-Haldane-FermN}(a).

Similar to the Shockley model, the time-dependent model parameter in the Haldane model can be simplified as
\begin{align}
 d_{\textbf{q}_c}^z(t)=\left\{
  \begin{array}{cc}
    u(t)-3\sqrt{3}t_2 &(\textbf{q}_c=\textbf{q}_{1/3/5}),\\
    u(t)+ 3\sqrt{3}t_2&(\textbf{q}_c=\textbf{q}_{2/4/6}),
  \end{array}
  \right.
\end{align}
where $\textbf{q}=(q_x,q_y)$ is the two-dimensional wave number, and $\textbf{q}_i$'s ($i=1,2,\cdots,6$) are six corners of the first brillouin zone, as shown in Fig. \ref{plot-Haldane-FermN}(b). The pKZ scaling behavior is given by
\begin{align}
  \mathcal{N}^{(\text{o}/\text{e})}_\text{pKZ}=\frac{1}{\frac{1}{d^{(\text{o}/\text{e})}(t_f)}-\frac{1}{d^{(\text{o}/\text{e})}(t_i)}}\frac{2}{\sqrt{3}\pi\gamma\tau_Q},
\end{align}
where we have $d^{(\text{o})}(t)= d_{\textbf{q}_{1/3/5}}^z(t)$ and $d^{(\text{e})}(t)= d_{\textbf{q}_{2/4/6}}^z(t)$.
The KZ scaling behavior is given by $\mathcal{N}_{\text{KZ}}=\frac{1}{2\sqrt{3}\pi^2\tau_Q}$. Under the quench protocols I-III with the parameter $\phi=\pi/2$, we can observe the two scaling behaviors, $\mathcal{N}_\text{KZ}$ and $\mathcal{N}_\text{pKZ}^{(\text{o}/\text{e})}$, appearing together or separately, as listed in Table \ref{table-1} and illustrated in Fig. \ref{plot-Haldane-FermN}(c)-(f).

\begin{table}[t]
  \renewcommand{\arraystretch}{1.4} %%设置表格行高
  \caption{Protocols, LLDs, and the consequent scaling behaviors of the fermion density. For Shockley model, the topological number is replaced with the winding number, $T\rightarrow W$, while for Haldane model, it is replaced with Chern number, $T\rightarrow C$. There are two types of scaling behaviors, KZ and pseudo-KZ scaling behavior. For the Shockley model, we have $\mathcal{N}_1=\mathcal{N}_\text{pKZ}^{(0)}$ and $\mathcal{N}_2=\mathcal{N}_\text{pKZ}^{(\pi)}$. For the Haldane model, we have $\mathcal{N}_1=\mathcal{N}_\text{pKZ}^{(\text{o})}$ and $\mathcal{N}_2=\mathcal{N}_\text{pKZ}^{(\text{e})}$.}
  \centering
\begin{tabular}{p{0.25cm}<{\centering}  p{2.4cm}<{\centering}|p{1.05cm}<{\centering}|p{1.4cm}<{\centering}|p{2.8cm}<{\centering}}
		\hline
        \hline
        \multirow{2}{*}{\uppercase\expandafter{\romannumeral 1}} &  \multirow{2}{*}{\includegraphics[width=0.13\textwidth]{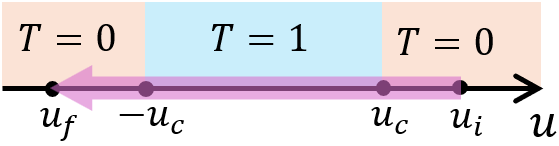}} & $\delta=\gamma$ & $2\mathcal{N}_\text{KZ} $ & Figs. \ref{plot-ShocM-FermN}(a) and \ref{plot-Haldane-FermN}(c)\\
        \cline{3-5}
         &   &   $\delta=-\gamma$  &-&-\\
        \hline
        \multirow{2}{*}{\uppercase\expandafter{\romannumeral 2}} &  \multirow{2}{*}{\includegraphics[width=0.13\textwidth]{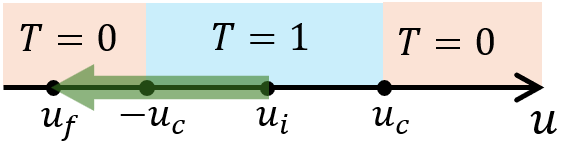}} & $\delta=\gamma$ & $\mathcal{N}_\text{KZ}$ &Figs. \ref{plot-ShocM-FermN}(b) and \ref{plot-Haldane-FermN}(d) \\
        \cline{3-5}
         &  &$\delta=-\gamma$ &$\mathcal{N}_1$& Figs. \ref{plot-ShocM-FermN}(c) and \ref{plot-Haldane-FermN}(e)\\
        \hline
        \multirow{2}{*}{\uppercase\expandafter{\romannumeral 3}} &  \multirow{2}{*}{\includegraphics[width=0.13\textwidth]{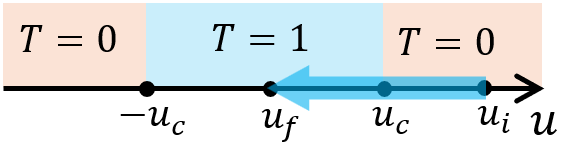}}  & $\delta=\gamma$ & $\mathcal{N}_\text{KZ}+\mathcal{N}_2$ & Figs. \ref{plot-ShocM-FermN}(d) and \ref{plot-Haldane-FermN}(f)\\
        \cline{3-5}
        & &$\delta=-\gamma$ &- &-\\
        \hline
        \hline
	\end{tabular}
\label{table-1}
\end{table}

\section{Discussion}\label{Sec-Discu}

At last, we address two important issues. First, from the rigorous solutions in our situations, we have the benefit of diagnosing the intertwined roles of the non-Hermitian terms, $\frac{1}{2}\{\rho,L_{s,j}^\dagger L_{s,j}\}$, and the quantum jump terms, $L_{s,j}\rho L_{s,j}^\dagger$, in the Lindblad equation in Eq. (\ref{Lindblad-RRM}). If we drop the quantum jump terms in Eq. (\ref{Lindblad-RRM}), the system would be described by an effective non-Hermitian Hamiltonian \cite{Yaohua_2023},
$H_\text{eff}=H-\frac{i}{2}\sum_{j}\sum_{s}L_{s,j}^\dagger L_{s,j}$. Qualitatively, this effective model can produce the KZ behavior in the excitation density, but no AKZ behavior can be observed in it. The quantum jump terms are responsible for the AKZ behavior alone. More detailed analysis shows the quantum jump terms also give a correction to the prefactor of the KZ scaling law to restore the full answer of the problem. While the fermion number $\mathcal{N}$ always contains no information about quantum jump, so we are not bothered by the AKZ behavior in measuring it.

Second, our conclusion is true for loss difference between the two sublattices of a bipartite lattice. Such kind of loss difference has been experimentally realized on the honeycomb photonic lattice \cite{Feng_2023}, in which a non-Hermitian Hamiltonian similar to ours has been discussed. In practice, the loss difference can also be realized for the two spin states, say $|\uparrow\rangle$ and $|\downarrow\rangle$, in the cold atom experiment \cite{Ren_2022}. With these state-of-the-art experimental progresses, we expect the KZ scaling law can be observed in near future.

\section{ACKNOWLEDGMENTS}

We thank Jian-Song Pan, Yan He, Adolfo del Campo, and Uwe R. Fischer for insightful discussions.  This work is supported by NSFC under Grant No. 11074177.

\appendix

\section{Two-Band Models in the Presence of Loss  } \label{appsec-1}

\subsection{Lindblad Master Equation}

The two-band model on a bipartite lattice in momentum space reads
\begin{align}\label{supM-H1}
  H=\sum_q \left(c_{a,q}^\dagger~ c_{b,q}^\dagger\right)\left[d_q^x\sigma_x+d_q^y\sigma_y+d_q^z\sigma_z\right]\binom{c_{a,q}}{c_{b,q}},
\end{align}
where $\sigma_\mu$'s represent a Pauli matrix, $c_{a,q}$ and $c_{b,q}$ denote the fermion operators on $a$ and $b$ sublattices respectively. By the canonical Bogoliubov transformation,
\begin{equation}\label{App-eta}
    \eta_{1,q}=u_{q}c_{a,q}+v_{q}c_{b,q}, ~~ \eta_{2, q}=-v_{q}^*c_{a,q}+u_{q}^*c_{b,q},
\end{equation}
with
\begin{equation}
  \left(u_{q}, v_{q}\right)=\left( \frac{\omega_{q}+d_q^z}{\sqrt{2\omega_{q}(\omega_{q}+d_q^z)}},
   \frac{d_q^x-id_q^y}{\sqrt{2\omega_{q}(\omega_{q}+d_q^z)}} \right),
\end{equation}
we obtain the diagonalized Hamiltonian,
\begin{align}\label{App-diagon-H0}
  H=\sum_{q}\omega_{q}(\eta_{1,q}^\dagger \eta_{1,q}-\eta_{2,q}^\dagger \eta_{2,q}),
\end{align}
where the quasiparticle dispersion reads $\omega_{q}=\sqrt{|d_q^x|^2+|d_q^y|^2+|d_q^z|^2}$. We suppose a quantum phase transition occurs for a specific model at the gapless point of the dispersion, $\omega_{q_c}=0$, with the critical mode $q_c$.

We introduce a linear ramp in the Hamiltonian in Eq. (\ref{supM-H1}),
\begin{align}\label{SupM-timePara}
  d_q^z~\rightarrow~d_q^z(t)=d_q^z(t_i)-\frac{t}{\tau_Q}, ~(t_i\leq t\leq t_f)
\end{align}
where $\tau_Q$ is the quench time, $t_i=0$, $t_f\gg0$, $d_q^z(t_i)\gg 0$, and $d_q^z(t_f)\ll 0$. We can rewrite the time-dependent Hamiltonian by a $4\times 4$ matrix,
\begin{equation}\label{app-mart-H}
  H(t)=\sum_{q}\phi_{q}^\dagger\left(
  \begin{array}{cccc}
    0 & 0& 0& 0 \\
    0&d_q^z(t_i)-t/\tau_Q &d^x_q-i d^y_q&0\\
    0&d^x_q+i d^y_q &-d_q^z(t_i)+t/\tau_Q&0 \\
    0 & 0&0 & 0
  \end{array}
  \right)\phi_{q}
\end{equation}
in the basis vector $\phi_{q}^\dagger=(|0\rangle, c_{a,q}^\dagger|0\rangle, c_{b,q}^\dagger|0\rangle,c_{a,q}^\dagger c_{b,q}^\dagger|0\rangle)$. The density matrix $\rho(t)$ can be expressed in the Fourier space as $\rho(t)=\bigotimes_{q}\rho_{q}(t)$, where
\begin{equation}\label{app-mart-rhoq}
\rho_{q}=\left(
\begin{array}{cccc}
  \rho_q^{(11)}&0&0&0\\
    0&\rho_q^{(22)}&\rho_q^{(23)}&0\\
0&\rho_q^{(32)}&\rho_q^{(33)}&0\\
   0 &0&0&\rho_q^{(44)}
\end{array}
\right).
\end{equation}
The matrix representations of the operators are given by
\begin{align}\label{app-mart-fer}
  &c_{a,q}=\left(
  \begin{array}{cccc}
    0&1&0&0 \\
    0&0&0&0 \\
    0&0&0&1 \\
    0&0&0&0
  \end{array}\right),~
  c_{b,q}=\left(
  \begin{array}{cccc}
    0&0&1&0 \\
    0&0&0&-1 \\
    0&0&0&0 \\
    0&0&0&0
  \end{array}\right),\\
  &\eta_{1,q}=\left(
  \begin{array}{cccc}
    0&u_q&v_{q}&0 \\
    0&0&0&v_{q} \\
    0&0&0&u_{q} \\
    0&0&0&0
  \end{array}\right),~
  \eta_{2,q}=\left(
  \begin{array}{cccc}
    0&-v_{q}^*&u_{q}^*&0 \\
    0&0&0&-u_{q}^* \\
    0&0&0&-v_{q}^* \\
    0&0&0&0
  \end{array}\right).
\end{align}

When the system interacts with the environment, the dynamics of the system can be described by the Lindblad master equation \cite{Dutta_2018, Vicari_2020, Campo_2020},
\begin{equation}\label{SupM-lindblad}
   \frac{\partial}{\partial t}\rho=-i[H(t),\rho]+\sum_{q}\mathcal{D}_q[\rho_q],
\end{equation}
where the term of loss is given by
\begin{align}\label{SupM-lossdis}
  \mathcal{D}_{q}[\rho_{q}]=&\sum_{s=a,b}\gamma_s\left(c_{s,q}\rho_{q}c_{s,q}^\dagger-
  \frac{1}{2}\{\rho_{q},c_{s,q}^\dagger c_{s,q}\}\right).
\end{align}

By Eqs. (\ref{app-mart-H}), (\ref{app-mart-rhoq}) and (\ref{app-mart-fer}), we get a set of first-order differential equations,
\begin{subequations}\label{App-Lindblad-set1}
  \begin{eqnarray}
   \frac{d\rho_q^{(11)}}{dt}&=& \gamma_a \rho_q^{(22)}+\gamma_b \rho_q^{(33)},\\
   \frac{d\rho_q^{(22)}}{dt}&=&-\gamma_a \rho_q^{(22)}+\gamma_b \rho_q^{(44)} +(i \Delta_q \rho_q^{(23)}-h.c.),~~~~~~\\
   \frac{d\rho_q^{(33)}}{dt}&=&-\gamma_b \rho_q^{(33)}+\gamma_a \rho_q^{(44)} -(i \Delta_q \rho_q^{(23)}-h.c.),~~~~~~\\
   \frac{d\rho_q^{(44)}}{dt}&=&-\gamma_a \rho_q^{(44)}-\gamma_b \rho_q^{(44)}, \label{App-Lindblad-set1-d}\\
   \frac{d\rho_q^{(23)}}{dt}&=&-\frac{\gamma}{2}\rho_q^{(23)} -2id_q^z(t) \rho_q^{(23)}-i\Delta_q^*R_q, \\
   \frac{d\rho_q^{(32)}}{dt}&=&-\frac{\gamma}{2}\rho_q^{(32)} +2id_q^z(t) \rho_q^{(32)}+i\Delta_q  R_q,
  \end{eqnarray}
\end{subequations}
where $\Delta_q=d^x_q+id^y_q$. The initial conditions at $t=t_i=0$ read
\begin{align}
  \rho_q^{(33)}=1, %\rho_q^{(11)}=\rho_q^{(22)}=\rho_q^{(44)}=0,\rho_q^{(23)}=(\rho_q^{(32)})^*=0
\end{align}
and other terms are $\approx 0$ for $d_q^z(t_i)\gg 0$.

By defining
\begin{align}\label{supM-definedR}
  &R_q=\rho_q^{(33)}-\rho_q^{(22)}=e^{-\frac{1}{2}\gamma t}\tilde{R}_q,~\rho_q^{(23)}=e^{-\frac{1}{2}\gamma t}\tilde{\rho}_q^{(23)},\nonumber\\
  &R_q'=\rho_q^{(11)}-\rho_q^{(44)}=e^{-\frac{1}{2}\gamma t}\tilde{R}_q',
\end{align}
we can transform the equations in Eq. (\ref{App-Lindblad-set1}) into a set of integral-differential equations
\begin{subequations}
\begin{align}
  &\tilde{R}_q'=(e^{\gamma t/2}-1)-\frac{\gamma_a-\gamma_b}{2}\int_{0}^{t}dt' \tilde{R}_q(t'), \label{App1-tildeR1R4} \\
  &\tilde{\rho}_q^{(23)}=-i\Delta_q\int_{0}^{t}\tilde{R}_q(t')e^{if(t,t')}dt' , \label{App1-tildeR23} \\
  &\frac{d\tilde{R}_q(t)}{dt}=\frac{\gamma_a-\gamma_b}{2}+\frac{(\gamma_a-\gamma_b)^2}{4}\int_{0}^{t}dt'\tilde{R}_q(t')\nonumber\\
&~~~-4|\Delta_q|^2\int_{0}^{t}\tilde{R}_q(t')\cos f(t,t')dt'.  \label{App1-tildeR}
\end{align}
\end{subequations}
where $f(t',t'')=2\int_{0}^{t}d_q^z(t')dt'-2\int_{0}^{t'}d_q^z(t'')dt''$

\subsection{Exact Solution in the Absence of Loss, $\gamma_a=\gamma_b=0$}
Now, we revisit the conventional Kibbke-Zurek mechanism (KZM) by the Lindblad formalism for $\forall\gamma_i=0$. Eq. (\ref{App1-tildeR}) becomes
\begin{align}\label{App-equR0}
  \frac{d R_q(s)}{ds}&=-4|\Delta_q|^2\tau_Q\int_{s_i}^{s}R_q(s')\cos\left(s^2-s'^2\right)ds'
\end{align}
where $s=[t-d_q^z(t_i)\tau_Q]/\sqrt{\tau_Q}$,  $s_i=-\sqrt{\tau_Q}d_q^z(t_i)$, and the initial condition $R_q(0)=1$.

According to KZM, only the modes near $q_c$ can be significantly excited ($|q- q_c|\sim 0$). For modes far away from $q_c$, we get the solution of Eq. (\ref{App-equR0}),
\begin{align}\label{App-R0-largeq}
  R_q=2|u_{q}|^2-1 ~~~(|q-q_c|\gg 0).
\end{align}
While in the vicinity of $q_c$ and in the absence of dissipation, the integral-differential equation can be solved iteratively as \cite{Kholodenko_2012, Fai_2013},
\begin{eqnarray}\label{Fresnel}
    R_q=1+\sum_{j=1}^{\infty}&~&(-4|\Delta_q|^2\tau_Q)^j\left[\prod_{l=1}^{j}\int_{s_i}^{s_{2l-1}}ds_{2l}\right.\nonumber\\
    &~ &\left.\int_{s_i}^{s_{2l}}ds_{2l+1}\cos(s_{2l}^2-s_{2l+1}^2)\right].
\end{eqnarray}\\
For slow ramp (i.e. $\tau_Q$ is large enough),  we can set $s_i\rightarrow-\infty$, which leads to the long-time asymptotic solution exactly,
\begin{eqnarray}\label{App-R0-qc}
  R_q=1+\sum_{j=1}^{\infty}\frac{(-4|\Delta_q|^{2}\tau_Q)^j\pi^j}{2^{2j-1}j!}=2 e^{-\pi\tau_Q|\Delta_q|^{2}}-1,~~~~~~
\end{eqnarray}
for $|q- q_c|\sim 0$. Usually we assume the linear behavior, $|\Delta_q|\propto |q-q_c|$ in Eq. (\ref{App-R0-qc}). Combining Eqs. (\ref{App-R0-largeq}) and (\ref{App-R0-qc}), we obtain the solution for $R_q$ after quench
\begin{align}
  R_q=2 e^{-\pi\tau_Q|\Delta_q|^{2}}+2|u_{q}|^2-1.
\end{align}
Near $q_c$, we have $(u_q, v_q)=(0,1)$ since $d_q^x\sim 0$, $d_q^y\sim0$, $d_q^z(t_f)\ll 0$ and
\begin{eqnarray}
  \left|\frac{u_q}{v_q}\right|&=&\frac{|d_q^z(t_f)|\sqrt{1+[(d_q^x)^2+(d_q^y)^2]/(d_q^z)^2}-|d_q^z(t_f)|}{\sqrt{(d_q^x)^2+(d_q^y)^2}}\nonumber\\
  &\approx& \frac{\sqrt{(d_q^x)^2+(d_q^y)^2}}{2|d_q^z(t_f)|}\approx0.
\end{eqnarray}
Thus Eq. (\ref{App-R0-qc}) is recovered when $q\rightarrow q_c$. Away from $q_c$, we have $e^{-\pi\tau_Q|\Delta_q|^{2}}=0$, so that Eq. (\ref{App-R0-largeq}) is recovered.

By $\rho_q^{(22)}+\rho_q^{(33)}=1$ and $\rho_q^{(33)}-\rho_q^{(22)}=R_q$, we get
\begin{eqnarray}
  \rho_q^{(22)} &=& (1-R_q)/2= 1- |u_{q}|^2-e^{-\pi\tau_Q|\Delta_q|^{2}} \\
  \rho_q^{(33)} &=& (1+R_q)/2=|u_{q}|^2+e^{-\pi\tau_Q|\Delta_q|^{2}},\\
  \rho_q^{(23)} &=&-\sqrt{e^{-\pi\tau_Q|\Delta_q|^{2}}(1-e^{-\pi\tau_Q|\Delta_q|^{2}})}e^{i\theta_{q}}-u_{q}^*v_{q},~~~~~~
\end{eqnarray}
and $\rho_q^{(32)}=(\rho_q^{(23)})^*$, where $\theta_{q}$ is the dynamical phase \cite{Dziarmaga_2010}.

\subsection{Exact Solution in the Presence of Uniform Loss, $\gamma_a=\gamma_b\neq0$}

For the situation of uniform loss, $\gamma_a=\gamma_b\neq0$, the equation in Eq. (\ref{App1-tildeR}) can be simplified to
\begin{eqnarray}\label{App-tildeR0}
  \frac{d \tilde{R}_q(s)}{ds}&=-4|\Delta_q|^2\tau_Q\int_{s_i}^{s}\tilde{R}_q(s')\cos\left(s^2-s'^2\right)ds',~~~~
\end{eqnarray}
where $s=[t-d_q^z(t_i)\tau_Q]/\sqrt{\tau_Q}$ and $s_i=-d_q^z(t_i)\sqrt{\tau_Q}$, which has the same form as that in Eq. (\ref{App-equR0}) and shares the same initial condition, $\tilde{R}_q(0)=R_q(0)=1$. The full solution of $\tilde{R}_q(t)$ is expressed by the Fresnel’s integrals in Eq. (\ref{Fresnel}). However, we only need to concern the solution at the final time ($t=t_f$), which can be exactly worked out by asymptotic analysis \cite{Kholodenko_2012, Fai_2013},
\begin{eqnarray}\label{Rtf}
    \tilde{R}_q(t_f)    &=&2 e^{-\pi\tau_Q|\Delta_q|^{2}}+2|u_{q}|^2-1,
\end{eqnarray}
in which we have defined $\bar{u}=[d_q^z(t_i)-d_q^z(t_f)]/2$. Moveover, the solution of Eq. (\ref{App1-tildeR1R4}) is given by
\begin{align}\label{supM-solubalR1R4}
  \tilde{R}_q'(t)=e^{\gamma t/2}\left(\rho_q^{(11)}-\rho_q^{(44)}\right)=e^{\gamma t/2}-1.
\end{align}
And by Eq. (\ref{App-Lindblad-set1-d}), we have $\rho_q^{(44)}=0.$

So, the solution of the set of equations in Eq. (\ref{App-Lindblad-set1}) is given by
\begin{subequations}\label{SupM-solu-bala}
\begin{align}
  &\rho_q^{(11)}=1-e^{-\bar{u}\gamma \tau_Q},\\
  &\rho_q^{(22)}=e^{-\bar{u}\gamma \tau_Q}\left(|v_q|^2-e^{-\pi\tau_Q|\Delta_q|^2}\right),\\
  &\rho_q^{(33)}=e^{-\bar{u}\gamma \tau_Q}\left(|u_q|^2+e^{-\pi\tau_Q|\Delta_q|^2}\right),\\
  &\rho_q^{(44)}=0,\\
  &\rho_q^{(23)}=\frac{-1}{e^{\bar{u}\gamma \tau_Q}}\left[\sqrt{e^{-\pi\tau_Q|\Delta_q|^{2}}(1-e^{-\pi\tau_Q|\Delta_q|^{2}})}e^{i\theta_{q}}
  +u_{q}^*v_{q}\right],\\
  &\rho_q^{(32)}=(\rho_q^{(23)})^*
\end{align}
\end{subequations}
at the final time $t_f$ of the quench.

\section{Analysis of Liouvillian quench dynamics in the dissipative two-level system } \label{App_LiouS}

By vectorized the density matrix of the dissipative two-level system, $\rho_q\rightarrow|\rho_q\rangle\!\rangle$, the Lindblad equation is given by
\begin{align}
  \frac{d|\rho_q\rangle\!\rangle}{dq}=\mathcal{L}_q|\rho_q\rangle\!\rangle,
\end{align}
where
\begin{align}
  |\rho_q\rangle\!\rangle&=\left. (|\rho_q^{(11)}\rangle\!\rangle,|\rho_q^{(22)}\rangle\!\rangle, \right.\nonumber\\
  & \left. |\rho_q^{(33)}\rangle\!\rangle, |\rho_q^{(44)}\rangle\!\rangle, |\rho_q^{(23)}\rangle\!\rangle, |\rho_q^{(32)}\rangle\!\rangle\right. )^T,
\end{align}
and $\mathcal{L}_q$ is superoperator which is given by
\begin{align}
\mathcal{L}_q&=\left(
\begin{array}{cccccc}
0  & \gamma_a   & \gamma_b & 0 & 0 &  0\\
0  & -\gamma_a  & 0        & \gamma_b & i\Delta_q & -i\Delta_q^*\\
0  & 0  & -\gamma_b        & \gamma_a & -i\Delta_q & i\Delta_q^*\\
0  & 0  & 0 & -\gamma & 0 &  0\\
0  & i\Delta_q^*  & -i\Delta_q^* & 0 & -2id_q^z-\frac{\gamma}{2} &  0\\
0  & -i\Delta_q  & i\Delta_q & 0 & 0 &  2id_q^z-\frac{\gamma}{2}\\
\end{array}
\right).
\end{align}
Here, we only consider the six nonzero elements of $\rho_q$, so $|\rho_q\rangle$ and $\mathcal{L}_q$ are expressed as a $6\times 1$ matrix and a $6\times6$ matrix, respectively.

We introduce a transfer matrix $T_q$ to diagonalize $\mathcal{L}_q$,
\begin{align}\label{App-diagLD}
  \frac{d |\tilde{\rho}_q\rangle\!\rangle}{dq}=T_q\mathcal{L}_qT_q^{-1}|\tilde{\rho}_q\rangle\!\rangle=\Lambda_q|\tilde{\rho}_q\rangle\!\rangle,
\end{align}
where
\begin{align}
  |\tilde{\rho}_q\rangle\!\rangle=T_q|\rho_q\rangle\!\rangle &=\left.(
   |\tilde{\rho}_q^{(3)}\rangle\!\rangle, |\tilde{\rho}_q^{(2,-)}\rangle\!\rangle, |\tilde{\rho}_q^{(2,+)}\rangle\!\rangle,  \right. \nonumber\\
  & \left.|\tilde{\rho}_q^{(1,-)}\rangle\!\rangle, |\tilde{\rho}_q^{(1,+)}\rangle\!\rangle, |\tilde{\rho}_q^{(0)}\rangle\!\rangle\right. )^\text{T},
\end{align}
and $\Lambda_q=\text{diag}\left(\lambda_3,\lambda_{2,-},\lambda_{2,+},\lambda_{1,-},\lambda_{1,+},\lambda_0\right)$ is the eigenvalue of $\mathcal{L}_q$. The expression of $\Lambda_q$ is shown in Eq. (\ref{Lq-lamdba}).

When $\Delta_q$ is near zero, the transfer matrix $T_q$ is approximately time independent,
\begin{align}
  T_q \approx
  \left(
  \begin{array}{cccccc}
    0 &0 &0 &2& 0 &0\\
    0 &0 &0 &0& 0 &1\\
    0 &0 &0 &0& 1 &0\\
    0 &-\sqrt{2} &0 &-\sqrt{2}& 0 &0\\
    0 &0 &-\sqrt{2} &-\sqrt{2}& 0 &0\\
    1 &1 &1 &1& 0 &0\\
  \end{array}
  \right)+O(\Delta_q)~~(\delta>0)
\end{align}
or
\begin{align}
  T_q \approx
  \left(
  \begin{array}{cccccc}
    0 &0 &0 &2& 0 &0\\
    0 &0 &0 &0& 1 &0\\
    0 &0 &0 &0& 0 &1\\
    0 &0 &-\sqrt{2} &-\sqrt{2}& 0 &0\\
    0 &-\sqrt{2} &0 &-\sqrt{2}& 0 &0\\
    1 &1 &1 &1& 0 &0\\
  \end{array}
  \right)+O(\Delta_q)~~(\delta<0)
\end{align}
So, even when the quench protocol outlined in Eq. (\ref{quenchprotocol}) is applied, the diagonalization of Eq. (\ref{App-diagLD}) remain valid for $\Delta_q\sim 0$.

Here, we assume that Hamiltonian only contain one critical point, and the parameter $d_q^z(t)$ can be simplified as $u(t)$.
The initial conditions read
\begin{align}
  |\tilde{\rho}_q(0)\rangle\!\rangle=\left\{
  \begin{array}{cc}
    (0,0,0,0,-\sqrt{2},1)^{\text{T}} & (u_i\delta>0),\\
    (0,0,0,-\sqrt{2},0,1)^{\text{T}}, & (u_i\delta<0),\\
  \end{array}
  \right.
\end{align}
Because Eq. (\ref{commutatorLS}) demonstrates that, for $\Delta_q\sim 0$, the commutator of the Liouvillian superoperators at different times is approximately zero, we can obtain the solution of Liouvillian dynamics under the quench protocol,
\begin{align}
  |\tilde{\rho}_q(t)\rangle\!\rangle=e^{\int_{0}^{t}\Lambda_q(t')dt'}|\tilde{\rho}_q(0)\rangle\!\rangle,
\end{align}
In the long-time evolution, i.e. $\gamma t_f\propto\gamma\tau_Q\gg 1$ and $e^{-\gamma t_f}\sim0$, the density matrix after vectorized reads
\begin{align}
  |\tilde{\rho}_q(t)\rangle\!\rangle\approx\left\{
  \begin{array}{cc}
    (0,0,0,0,-\sqrt{2}y,1)^{\text{T}} & (u_i\delta>0),\\
    (0,0,0,0,0,1)^{\text{T}} & ( u_i\delta<0),\\
  \end{array}
  \right.
\end{align}
where
\begin{align}
  y=e^{\int_{0}^{t}\lambda_{1,+}(t')dt'}.
\end{align}

So, using the relation of $|\rho_q (t)\rangle\!\rangle=T^{-1}|\tilde{\rho}_q(t)\rangle\!\rangle$, we can write the solution as
\begin{eqnarray}
  |\rho_q(t)\rangle\!\rangle&=&\left\{
  \begin{array}{cc}
    (1-y,0,y,0,0,0)^{\text{T}} & (u_i, \delta>0),\\
    (1-y,y,0,0,0,0)^{\text{T}} & (u_i, \delta<0),\\
    (1,0,0,0,0,0)^{\text{T}}   & (u_i\delta<0).
  \end{array}
  \right.~~~~~
  \end{eqnarray}
For the case with $u_i\delta>0$, we write the operator form of density matrix,
\begin{subequations}\label{solu_LiouQD}
\begin{align}
\rho_q^{(11)}(t)&=1-e^{\int_{0}^{t}\lambda_{1,+}(t')dt'},\\
  \rho_q^{(22)}(t)&=\frac{1-\text{sgn}(\delta)}{2}e^{\int_{0}^{t}\lambda_{1,+}(t')dt'} ,\\
  \rho_q^{(33)}(t)&=\frac{1+\text{sgn}(\delta)}{2}e^{\int_{0}^{t}\lambda_{1,+}(t')dt'},
\end{align}
\end{subequations}
and other terms are zero in the long-time evolution.

\section{Rigorous Solution in the Presence of Loss Difference, $\gamma_a\neq \gamma_b$, in the Rice-Mele model}
\label{appsec-3}

\begin{widetext}
\begin{table*}[t]
	\renewcommand{\arraystretch}{1.7} %%设置表格行高
	\centering
	\caption{Excitation and fermion densities in the Rice-Mele model for $\gamma_a=\gamma_b=0$, $\gamma_a=\gamma_b\neq0$, and $\gamma_a\neq\gamma_b$.}
	\begin{tabular}{p{3.7cm}<{\centering}|p{2.7cm}<{\centering}|p{3.7cm}<{\centering}|p{6cm}<{\centering}}
		\hline
		\hline
           & $\gamma_a=\gamma_b=0$ & $\gamma_a=\gamma_b\neq0$&$\gamma_a\neq\gamma_b ~~(\delta=\gamma_a-\gamma_b\neq0)$\\
        \hline
        Excitation density $n=\frac{1}{N}\sum_{q}p_q$ & $\frac{1}{2\pi\sqrt{\tau_Q}}$ & $\frac{e^{-\bar{u}\gamma\tau_Q}}{2\pi\sqrt{\tau_Q}}+\frac{1}{2}(1-e^{-\bar{u}\gamma\tau_Q})$  &
        $\frac{e^{-\bar{u}\gamma\tau_Q}}{2\pi\sqrt{\tau_Q}}+\frac{1}{2}(1-e^{-\bar{u}\gamma\tau_Q})+\frac{e^{\bar{u}\delta\tau_Q}-1}{ e^{\bar{u}\gamma\tau_Q}}\frac{1}{4\pi\sqrt{\tau_Q}}$ \\
        \hline
        Fermion number at $a$ $\mathcal{N}_a=\frac{1}{N}\sum_{q}\text{Tr}(\rho c_{a,q}^\dagger c_{a,q})$ & $\mathcal{N}_a^\text{GS}-\frac{1}{2\pi\sqrt{\tau_Q}}$ & $e^{-\bar{u}\gamma\tau_Q}\left(\mathcal{N}_a^\text{GS}-\frac{1}{2\pi\sqrt{\tau_Q}}\right)$ &
        $e^{-\bar{u}\gamma\tau_Q}\left(\mathcal{N}_a^\text{GS}-\frac{1}{2\pi\sqrt{\tau_Q}}\right)$ \\
        \hline
        Fermion number at $b$ $\mathcal{N}_b=\frac{1}{N}\sum_{q}\text{Tr}(\rho c_{b,q}^\dagger c_{b,q})$ & $\mathcal{N}_b^\text{GS}+\frac{1}{2\pi\sqrt{\tau_Q}}$ & $e^{-\bar{u}\gamma\tau_Q}\left(\mathcal{N}_b^\text{GS}+\frac{1}{2\pi\sqrt{\tau_Q}}\right)$ &
        $e^{-\bar{u}\gamma\tau_Q}\left(\mathcal{N}_b^\text{GS}+\frac{1}{2\pi\sqrt{\tau_Q}}\right)+\frac{e^{\bar{u}\delta\tau_Q}-1}{ e^{\bar{u}\gamma\tau_Q}}\frac{1}{2\pi\sqrt{\tau_Q}}$\\
        \hline
	\end{tabular}
\label{supM-list}
\end{table*}
\end{widetext}

\begin{figure}[t]%!htbp]
  \begin{center}
		\includegraphics[width=2.8 in,angle=0]{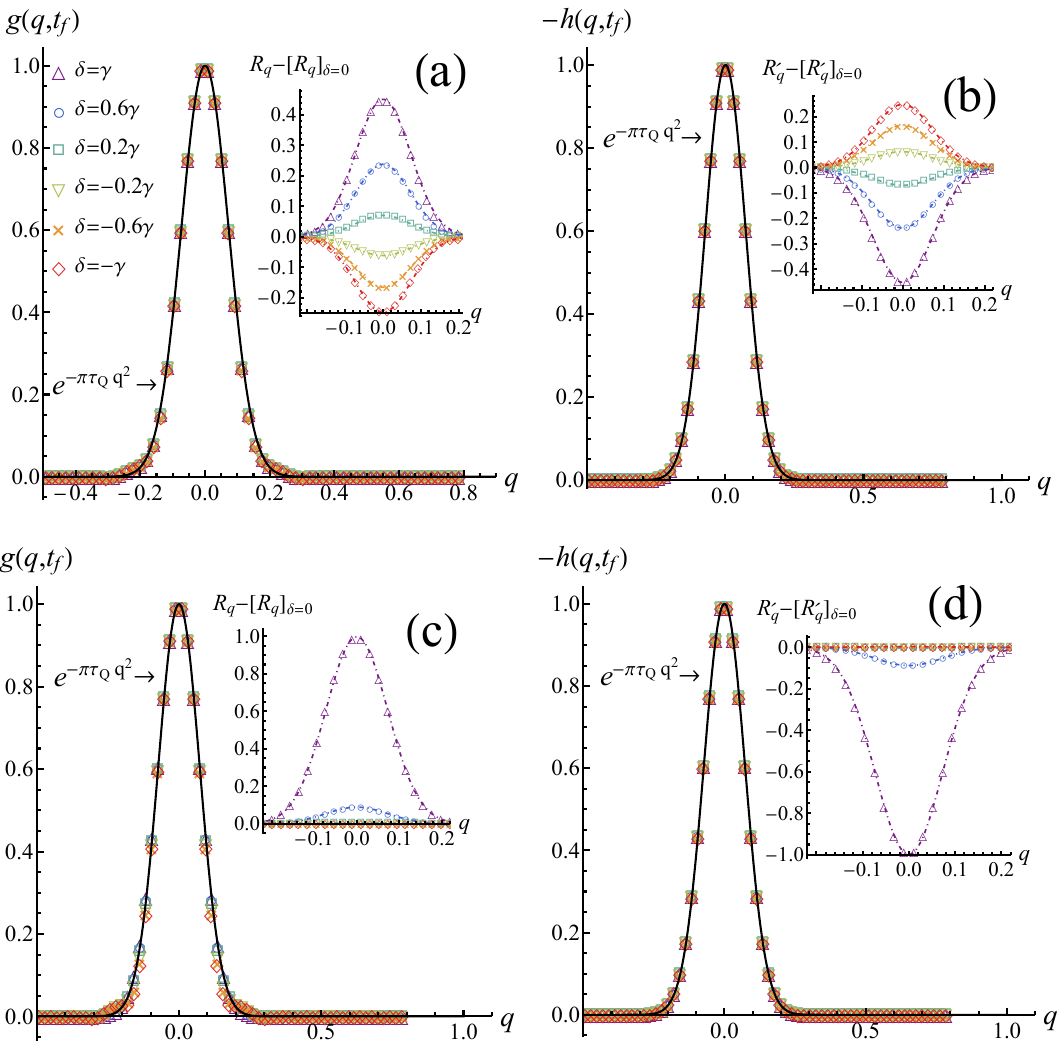}
  \end{center}
  \caption{The functions $g(q,t_f)$ and $h(q,t_f)$ versus $q$ in the Rice-Mele model in the presence of loss difference. We have fixed the parameters, $u_i=2$, $u_f=-2$, $\tau_Q=30$, and $\gamma_a+\gamma_b=0.01$ for panels (a) and (b), while $\gamma_a+\gamma_b=0.1$ for panels (c) and (d). The numerical results indicated by the colored dingbats are obtained from solving the equations in Eq. (\ref{App-Lindblad-set1}). The analytical functions are denoted by the black lines. In the insets of panels (a) and (c), the dash-dotted lines represent the solution in Eq. (\ref{supm-solR}). While in the insets of panels (b) and (d), the dash-dotted lines represent the solution in Eq. (\ref{supm-solR1R4}). We see the analytical expressions match the numerical results rigorously.}
  \label{plot-SupM-gt}
\end{figure}

We now discuss the loss difference, i.e. $\delta\equiv\gamma_a-\gamma_b\neq 0$ in the Rice-Mele model. The equations in Eqs. (\ref{App1-tildeR1R4}) and (\ref{App1-tildeR}) become
\begin{subequations}
\begin{eqnarray}\label{app-imb-tildeRp}
  \frac{d\tilde{R}_q'(t)}{dt}&=&e^{\gamma t/2}-1-\frac{\delta}{2}\int_{0}^{t}dt' \tilde{R}_q(t'),\\
  \frac{d\tilde{R}_q(t)}{dt}&=&\frac{\delta}{2}+\left(\frac{\delta}{2}\right)^2\int_{0}^{t}dt'\tilde{R}_q(t')\nonumber\\
  -&4&|\Delta_q|^2\int_{x_i}^{x}\tilde{R}_q(x')\cos\left[\frac{x^2-(x')^2}{\tau_Q}\right]dx',~~~ \label{app-imb-tildeR}
%\int_{0}^{t}\tilde{R}_q(t')\cos\left(2\int_{0}^{t}\epsilon(t')dt'-2\int_{0}^{t'}\epsilon(t'')dt''\right)dt'
\end{eqnarray}
\end{subequations}
where $x=t-u_i\tau_Q$ and $x_i=-u_i\tau_Q$. The initial conditions read $\tilde{R}_{q}(0)=1$ and $\tilde{R}'_{q}(0)=0$. We speculate the solution in two steps. First, let us consider the solution for $q=0$, which means $\Delta_q=0$ so that the last term in Eq. (\ref{app-imb-tildeR}) disappears and the solution is given by
\begin{equation}
  \tilde{R}_{0}(t)=e^{\delta t/2},~~~~\tilde{R}'_{0}(t)=e^{\gamma t/2}+e^{\delta t/2}-2.
\end{equation}
Next, we consider $q\neq0$ and speculate the solution of $\tilde{R}_{q}(t)$ as,
\begin{align}\label{gh-1}
\begin{array}{cc}
  \tilde{R}_{q}(t)&=\left[\tilde{R}_{q}(t)\right]_{\delta=0}+g(q,t)(e^{\delta t/2}-1),\\
  \tilde{R}'_{q}(t)&=\left[\tilde{R}'_{q}(t)\right]_{\delta=0}+h(q,t)(e^{\delta t/2}-1),
\end{array}
\end{align}
where $g(0,t)=h(0,t)=1$. For arbitrary time $t$, $[\tilde{R}'_{q}(t)]_{\delta=0}$ is just the solution in Eq. (\ref{supM-solubalR1R4}). While $[\tilde{R}_{q}(t)]_{\delta=0}$ should resort to the iterative method like that in Eq. (\ref{Fresnel}), but again we can only concern the final time of the quench and work out the asymptotic result as ($t_f=2\bar{u}\tau_Q$ here)
\begin{subequations}\label{supm-soltildeR}
\begin{eqnarray}
  \tilde{R}_{q}=\tilde{R}_{q}(t_f)&=&[\tilde{R}'_{q}(t_f)]_{\delta=0}+e^{-\pi\tau_Q q^2}(e^{\bar{u}\delta \tau_Q}-1),~~~~\\
  \tilde{R}_{q}'=\tilde{R}_{q}'(t_f)&=&e^{\bar{u}\gamma \tau_Q}-1-e^{-\pi\tau_Q q^2}\left(e^{\bar{u}\delta \tau_Q}-1\right).~~~~
\end{eqnarray}
\end{subequations}
In order to verify these analytical expressions, we have defined two functions according to Eq. (\ref{gh-1}),
\begin{align}\label{gh-2}
 \begin{array}{cc}
  g(q,t_f)&=\frac{\tilde{R}_{q}(t_f)-\left[\tilde{R}_{q}(t_f)\right]_{\delta=0}}{e^{\delta t_f/2}-1},\\
  h(q,t_f)&=\frac{\tilde{R}'_{q}(t_f)-\left[\tilde{R}'_{q}(t_f)\right]_{\delta=0}}{e^{\delta t_f/2}-1},
  \end{array}
\end{align}
and compare them with the numerical results by solving the equations in Eq. (\ref{App-Lindblad-set1}). As shown in Fig. \ref{plot-SupM-gt}, we see the analytical solution in Eq. (\ref{supm-soltildeR}) is quite satisfactory. Thus, we can list all solutions for the case of $\gamma_a\neq\gamma_b$ as below,
\begin{subequations}\label{app-imb-solu}
\begin{align}
  R_{q}=&\frac{2 e^{-\pi\tau_Q q^2}+2|u_q|^2-1}{e^{\bar{u}\gamma \tau_Q}}+e^{-\pi\tau_Q q^2}\frac{e^{\bar{u}\delta \tau_Q}-1}{e^{\bar{u}\gamma \tau_Q}},  \label{supm-solR} \\
  R_{q}'=&1-e^{-\bar{u}\gamma \tau_Q}-e^{-\pi\tau_Q q^2}\frac{e^{\bar{u}\delta \tau_Q}-1}{e^{\bar{u}\gamma \tau_Q}},  \label{supm-solR1R4} \\
  \rho^{(11)}_{q}=&1-e^{-\bar{u}\gamma \tau_Q}-e^{-\pi\tau_Q q^2}\frac{e^{\bar{u}\delta \tau_Q}-1}{e^{\bar{u}\gamma \tau_Q}},\\
  \rho^{(22)}_{q}=& e^{-\bar{u}\gamma \tau_Q}\left(1-|u_q|^2-e^{-\pi\tau_Q q^2}\right), \\
  \rho^{(33)}_{q}=&\frac{|u_q|^2+e^{-\pi\tau_Q q^2}}{e^{\bar{u}\gamma \tau_Q}}+e^{-\pi\tau_Q q^2}\frac{e^{\bar{u}\delta \tau_Q}-1}{e^{\bar{u}\gamma \tau_Q}},\\
  \rho^{(44)}_{q}=&0.
\end{align}
\end{subequations}

The density of excitations is worked out as
\begin{eqnarray}\label{supM-imbaln}
  n=\frac{e^{-\bar{u}\gamma\tau_Q}}{2\pi\sqrt{\tau_Q}}+\frac{1}{2}(1-e^{-\bar{u}\gamma\tau_Q})+\frac{e^{\bar{u}\delta\tau_Q}-1}{ e^{\bar{u}\gamma\tau_Q}}\frac{1}{4\pi\sqrt{\tau_Q}}.~~~~
\end{eqnarray}
In Eq. (\ref{supM-imbaln}), the first term comes from $\rho_q^{(22)}$ and  $\rho_q^{(33)}$ that is generated by the non-Hermitian terms of the Lindblad equation, the second term is attributed to $\rho_q^{(11)}$ that is generated by the quantum jump terms, and the third term comes from $\rho_q^{(11)}$ and $\rho_q^{(33)}$. In the limit of loss difference (LLD), i.e. $\gamma=\delta$, a universal KZ scaling law emerge as $n-1/2\approx n_\text{KZ}=\frac{1}{4\pi\sqrt{\tau_Q}}$ for large enough $\tau_Q$, in which $\rho_q^{(33)}$ produces $2n_{\text{KZ}}$ while $\rho_q^{(11)}$ generates $-n_{\text{KZ}}$.

The fermion numbers on sublattice $a$ and $b$ are worked out as
\begin{subequations}\label{App-imb-n}
\begin{align}
  \mathcal{N}_a=&e^{-\bar{u}\gamma \tau_Q}\left(\mathcal{N}_a^\text{GS}-\frac{1}{2\pi\sqrt{\tau_Q}}\right)   \label{App-imb-na}\\
  \mathcal{N}_b=&e^{-\bar{u}\gamma\tau_Q}\left(\mathcal{N}_b^\text{GS}+\frac{1}{2\pi\sqrt{\tau_Q}}\right)+\frac{e^{\bar{u}\delta\tau_Q}-1}{ e^{\bar{u}\gamma\tau_Q}}\frac{1}{2\pi\sqrt{\tau_Q}}.  \label{App-imb-nb}
\end{align}
\end{subequations}
respectively. $\mathcal{N}_{a/b}^\text{GS}$ represents the fermion density on the $a(b)$ sublattice of ground state. Finally, we summarize the above results in Table \ref{supM-list}.

\section{The Lindblad equation without the quantum jump term for the Rice-Mele Model} \label{app-Qjumpt}

By ignoring the quantum jump terms in Eq. (\ref{SupM-lindblad}), we get
\begin{equation}\label{SupM-lindblad-2}
   \frac{\partial}{\partial t}\rho=-i[H(t),\rho]-\frac{1}{2}\sum_{q}\sum_{s=a,b}(\gamma_s\{\rho_{q},c_{s,q}^\dagger c_{s,q}\}).
\end{equation}
By repeating the same steps as in Appendix \ref{appsec-3}, we have
\begin{eqnarray}
  n&=&\frac{1}{2N}\text{Tr}(\rho\eta_{1,q}^\dagger\eta_{1,q})+\frac{1}{2}\text{Tr}[\rho(1-\eta_{2,q}^\dagger\eta_{2,q})] \nonumber\\
  &=&\frac{e^{-\bar{u}\gamma\tau_Q}}{2\pi\sqrt{\tau_Q}}+\frac{e^{\bar{u}\delta\tau_Q}-1}{ e^{\bar{u}\gamma\tau_Q}}\frac{1}{2\pi\sqrt{\tau_Q}},
\end{eqnarray}
So we see the AKZ behavior disappears and the KZ scaling law, $n_{\text{KZ}}\sim\frac{1}{2\pi\sqrt{\tau_Q}}$, is qualitatively produced at the LLD ($\delta=\gamma$). To quantitatively restore of the correct answer of KZ scaling law, $n_{\text{KZ}}=\frac{1}{4\pi\sqrt{\tau_Q}}$, the quantum jump terms must be included.

%%\nocite{*}
%\bibliographystyle{apsrev4-2}%aapmrev4-2}%plain%IEEEtran%ieeetr% unsrt
%\bibliography{citation}

\providecommand{\noopsort}[1]{}\providecommand{\singleletter}[1]{#1}%

\end{document}